\documentclass[12pt]{article}
\usepackage{graphicx,amsmath,amssymb,color,bm}
\usepackage{amsmath,amssymb,amsthm,color}
\usepackage{graphicx}
\usepackage{cite}
\usepackage{epsfig}
\usepackage{lineno}
\renewcommand{\baselinestretch}{2}

\begin{document}

\centerline {\Large\textbf {Magnetic quantization in multilayer graphenes  }}

\centerline{ Chiun-Yan Lin$^{1}$, Jhao-Ying Wu$^{1\ddag}$, Yih-Jon Ou$^{1}$, Yu-Huang Chiu$^{2\dag}$, Ming-Fa Lin$^{1}$$^{*}$}
\centerline{$^1$ Department of Physics, National Cheng Kung University, Tainan, Taiwan}
\centerline{$^2$ National center for theoretical science, Taiwan}

\vskip0.6 truecm
\noindent

\begin{abstract}

Essential properties of multilayer graphenes are diversified by the number of layers and the stacking configurations. For an $N$-layer system, Landau levels are divided into $N$ groups, with each identified by a dominant sublattice associated with the stacking configuration. We focus on the main characteristics of Landau levels, including the degeneracy, wave functions, quantum numbers, onset energies, field-dependent energy spectra, semiconductor-metal transitions, and crossing patterns, which are reflected in the magneto-optical spectroscopy, scanning tunneling spectroscopy, and quantum transport experiments. The Landau levels in AA-stacked graphene are responsible for multiple Dirac cones, while in AB-stacked graphene the Dirac properties depend on the number of graphene layers, and in ABC-stacked graphene the low-lying levels are related to surface states. The Landau-level mixing leads to anticrossings patterns in energy spectra, which are seen for intergroup Landau levels in AB-stacked graphene, while in particular, a formation of both intergroup and intragroup anticrossings is observed in ABC-stacked graphene.
The aforementioned magneto-electronic properties lead to diverse optical spectra, plasma spectra, and transport properties when the stacking order and the number of layers are varied. The calculations are in agreement with optical and transport experiments, and novel features that have not yet been verified experimentally are presented.
\end{abstract}

\vskip0.6 truecm

$\mathit{PACS}$: 81.05.U-, 78.67.Pt, 71.70.Di

\noindent{Contents}

\pagenumbering{Roman}

\addcontentsline{toc}{chapter}{摘要}

\vskip0.6 truecm


\vskip0.6 truecm

\noindent {1. Introduction} \hfill ............................................................................................................~~~01

\vskip0.5 truecm

\noindent

\noindent {2. The generalized tight-binding model}.........................................................................~~~08

\noindent {2.1 Monolayer graphene}
\hfill .................................................................................................~~~08

\noindent {2.2 Multilayer graphene}..................................................................................................~~~11

\noindent {2.2.1 AA stacking}
\hfill.........................................................................................................~~~11

\noindent {2.2.2 AB stacking}
\hfill.........................................................................................................~~~13

\noindent {2.3.3 ABC stacking}
\hfill......................................................................................................~~~15

\vskip0.5 truecm

\noindent {3. Magnetic quantization in monolayer graphene}..........................................................~~~17

\noindent {3.1 Dirac-cone band structure }
\hfill......................................................................................~~~17

\noindent {3.2 Landau levels and Landau wave functions }
\hfill................................................................~~~19

\noindent {3.3 Oscillating Landau subbands under a modulated magnetic field }
\hfill..............................~~~22

\noindent {3.4 Geometry-induced non-uniform magnetic field in curved graphene
systems}
\hfill.............~~~24

\vskip0.5 truecm

\noindent {4. Magnetic quantization in AA-stacked graphene}........................................................~~~26

\noindent {4.1 Multiple Dirac-cone band structure}
\hfill .........................................................................~~~26

\noindent {4.2 Landau levels and Landau wave functions}
\hfill..............................................................~~~27

\noindent {4.3 Magnetic-field-dependent oscillation of the Fermi level}
\hfill............................................~~~34

\noindent {4.4 Magneto absorption spectrum}
\hfill ..................................................................................~~~34

\vskip0.5 truecm

\noindent {5. Magnetic quantization in AB-stacked graphene}........................................................~~~36

\noindent {5.1 Band structure: hybridization with monolayer graphene}
\hfill .......................................~~~36

\noindent {5.2 Landau levels and Landau wave functions}
\hfill..............................................................~~~38

\noindent {5.3 Symmetry-breaking effect on Landau level spectrum}
\hfill..............................................~~~40

\noindent {5.4 Magneto absorption spectrum}
\hfill ................................................................................~~~45

\noindent {5.5 Magneto plasmons}
\hfill..................................................................................................~~~46

\vskip0.5 truecm

\noindent {6. Magnetic quantization in ABC-stacked graphene}.....................................................~~~47

\noindent {6.1 Band structure: surface-localized states and sombrero-shaped energy bands} \hfill ..........~~~48

\noindent {6.2 Landau levels and Landau wave functions}
\hfill................................................................~~~50

\noindent {6.3 Intra- and inter-group Landau-level anticrossings in the field-dependent energy spectrum}
\hfill.............................................................................................................................~~~52

\noindent {6.4 Peculiar quantization rule for sombrero-shaped energy bands}.................................~~~58

\noindent {6.4.1 Inverse dependence of Landau-level energies on the magnetic field}
\hfill......................~~~58

\noindent {6.4.2 Periodical quantization effect related to the inverse of the critical magnetic field}
\hfill......................................................................................................................................~~~61

\noindent {6.5 Magneto absorption spectrum}
\hfill................................................................................~~~61

\vskip0.5 truecm

\noindent {7. Differences among AA-, AB- and ABC-stacked graphenes}.......................................~~~62

\noindent {7.1 Landau-level spectrum}
\hfill ...........................................................................................~~~62

\noindent {7.2 Wave functions and density of states of Landau levels}
\hfill .............................................~~~64

\noindent {8. Concluding remarks}
\hfill.............................................................................................~~~66

\noindent {Acknowledgments}
\hfill.........................................................................................................~~~70

\noindent {References}
\hfill.....................................................................................................................~~~71

\noindent {Figure captions}
\hfill.............................................................................................................~~~94

\pagebreak
\renewcommand{\baselinestretch}{2}
\newpage

\vskip 0.6 truecm



\setcounter{page}{1}
\pagenumbering{arabic}

\newpage

\bigskip

\section{Introduction}

\bigskip

\bigskip

Graphene is a one-atom-thick material made up of carbon atoms arranged in a hexagonal lattice structure\cite{Science306;666,Proc102;10451}.
The particular electronic properties and the high carrier mobility (up to 15,000 cm$^{2}$V$^{-1}$s$^{-1}$) \cite{Science342;720,NatMat6;183,Nature438;197,PRL100;016602,NatTech3;206,SciRep4;4558,NatTech5;487}
make this new material a promising candidate for next-generation nano-devices\cite{NatTech2;605,NatCom3;906,NatTech9;780,NanoLett12;2773,NatPho4;611,
NanoLett10;4285}.
Graphene is the building block of other carbon allotropes, geometrically in terms of curling, cutting, folding and stacking$-$ such as the synthesized zero-dimensional (0D) fullerenes\cite{Nat385;780}, 1D graphene nanoribbons\cite{Science319;1229,Nature488;627,
NatTech5;727} and carbon nanotubes\cite{Nattech354;56}, 2D layered graphenes
\cite{NatMat6;183,Science306;666} and 3D bulk graphite\cite{ADVPHY30;139}. These low-dimensional systems have been a focus of intense interest in the areas of
electronic properties\cite{PRL101;046803,PRL103;256801,PRB84;161406,PRB87;165102,PRB46;4540,CARBON32;289,PRB81;165105,PRL97;036803,
PRB81;115315,Jnn11;4938,PRB84;075451,PRL98;206805,PRB84;125455,JPSJ76;024701,
PRB75;193402,PRB74;161403,PRB75;155115,PRB87;115422,PRB87;085424,PRB87;075417,
JAP114;233701,CPL550;104,PRB77;155416,PRL99;216802,JAP110;013725,JPSJ65;505,
PRB67;045405,PRB62;16092,Science304;1129,PRB59;8271,PRB54;17954,
Nanotechnology18;495401,SCIENCE312;1191,PRL107;086601,PRB83;165429,PRB78;205425,
Rep76;056503,JPCM18;5849,PRB84;165404,PRB82;035409,PRL73;245426,PRB83;220503,
PRB87;155116,RevModPhys81;109,APL97;101905,PhyE40;1722,PRL96;086805,PRB77;085426,Ann326;721,
Carbon42;2975,PRB84;205448,PRB83;165443,RSCAdv4;56552,PRB80;165409,PRB90;205434,Nature459;820,PRL102;256405,Science313;951,Nature7;948,
NatNanotechnol4;383,SCIENCE320;206},
optical properties\cite{SCIREP3;3143,PRL98;157402,ACSNano4;1465,PRB73;245411,PRL100;087403,PRL102;166401,
PRL102;037403,PRB79;115441,PRB78;235408,Nature4;532,PRB83;245418,PRL110;246803,PRL111;077402,
PRL97;266405,PRL98;197403,PRL101;267601,SynMet162;800,PRL100;151,PRB85;245410,
PRB77;115313,JAP108;043509,PRB83;125302,PRR89;045419,PRL104;176404,PRB73;144427,
Nature459;820,NJP15;015010,APL103;041907,CARBON69;151,APL97;101905,PhyE40;1722,PRB46;4540,CARBON32;289,PRB81;165105},
Coulomb excitations\cite{PRB74;085406,PLA352;446,PRB86;125434,PRB89;165407,ACSnano5;1026,
NanoResLett7;134,PRB85;235444,PRB80;085408,PRB84;035439,PRB84;115420,PRB75;115314,
JAP109;113721}
and quantum Hall transport properties\cite{PRL107;086601,Nanotechnology23;052001,PRB88;121302,PRL107;126806,PRB82;165404,
NanoLett13;1627,PRX2;011004,NatMater7;151,Nature438;197,
SCIENCE315;1379,PRL95;146801,Nature438;201,NatPhys2;177,NatPhys7;621,NatPhys7;953}.
It has been demonstrated in theoretical and experimental research studies that the essential properties are dominated by the interplay between the external fields and the specific geometric symmetries, and related to the geometric boundary conditions\cite{PRB59;8271,JPSJ65;505,PRB67;045405,
PRB62;16092,Science304;1129,PRB54;17954,Nanotechnology18;495401,SCIENCE312;1191,
PRL107;086601,PRL98;206805}.
The results are rather different from those in conventional metals and semiconductors\cite{JAP76;1363,RMP54;437}. To illustrate the geometry-dependent response under magnetic quantization, we thoroughly review multilayer graphenes with different stacking configurations.

Planar monolayer graphene is a starting point toward the broad investigation of other carbon allotropes with diverse geometric symmetries. It was a theoretically studied material until in 2004 a group led by Andre Geim and Kostya Novoselov extracted monolayer graphenes from bulk graphite by using the micro-mechanical exfoliation technique\cite{Science306;666}.
Their success in isolating graphene sheets immediately triggered much activity in obtaining high-quality large graphene sheets. As a result of the efforts of many groups over the past decade, the improved quality is adequate for research experiments and industry applications\cite{NatCom4;2096,Science342;720,NanoLett11;3612,NatTech5;574,
ACSNano6;8241,SCIENCE324;1312}. The electronic structures of graphenes were first derived by Wallace from the simple nearest-neighboring tight-binding model\cite{PR71;622}. The calculated results show that monolayer graphene is a gapless semiconductor with a vanishing density of states at the Fermi level ($E_{F}=0$). The low-lying energy dispersions linearly depend on the wave vector, with crossings of conduction and valence bands at the $K$ and $K^{'}$ points (the hexagonal corners of the first Brillouin zone).
This leads to a conical energy spectrum around the corners $K$ and $K^{'}$ of the Brillouin zone, resembling the Dirac fermion spectrum of relativistic quantum particles, with an effective speed of light $\simeq c/300$, where $c$ is the light speed. As a result, the quasi-particles in graphene can be described by a Dirac-like Hamiltonian\cite{Nature438;197,RevModPhys81;109,PRL96;086805}. The conical structures are known as Dirac cones, and the similarly band crossing points, $K$ and $K^{'}$, are known as Dirac points.

Graphene sheets are held together by weak Van der Waals interactions.
The stacked graphenes have been experimentally characterized by infrared spectroscopy\cite{PRL102;037403,Nature4;532,PRB78;235408,PRB79;115441}, transmission electron microscopy\cite{PRL102;015501,JChemPhys129;234709,JAP109;093523,PRB81;161410,
ACSNano6;5680},
scanning tunneling microscopy (STM) and spectroscopy (STS)\cite{PRL109;116802,PRB76;201402,NATURE6;109,NatMat12;887,PRL106;126802,
NanoLett3;5153,PRB77;155426,SurfSci610;53,SCIENCE324;924,NATURE467;185,
NATURE7;245,PRL109;176804,NATURE7;649},
Raman spectroscopy
\cite{PRL97;187401,NanoLett12;1749,NanoLett11;164,Carbon46;272,NanoLett12;5539,
PCCP9;1276, JAP110;013720}, and angle-resolved photoelectron spectroscopy (ARPES)\cite{Nature3;36,PRB77;155303,PRL103;226803,PPL98;206802,PRB88;075406,
PRB88;155439,PRL102;056808}.
The sacking orders are arranged in specified sequences, essentially AA\cite{PRL102;015501,JAP109;093523,JChemPhys129;234709,PRB81;161410}, AB\cite{PRB77;155426,PRL97;187401,NanoLett12;1749,NanoLett11;164,SurfSci610;53,
Carbon46;272,NanoLett12;5539,PRB81;161410}
and ABC stacking\cite{PRB77;155426,NanoLett11;164,SurfSci610;53,
Carbon46;272,NanoLett12;5539,ACSNano6;5680,PRB81;161410}.
A haphazard stacking of graphene sheets is called turbostratic structure\cite{PCCP9;1276,JAP110;013720}. AB and ABC stacking configurations are common stacking orders of natural graphite\cite{PRSLSA106;749,PRSLSA181;101}, while AA stacking configuration has been found only in intercalated graphite\cite{BSCF187;999}.
It was reported that natural graphite typically contains 80 $\%$ of volume fraction of AB configuration, 14 $\%$ of ABC configuration, and the rest part is about the other turbostratic structures\cite{PRSLSA106;749,PRSLSA181;101}. Many works on graphene synthesis have been devoted to obtain few-layer graphenes by exfoliation of highly orientated pyrolytic graphite (mechanical and chemical processes)\cite{Carbon66;340,NanoRes6;95,Science306;666},
chemical and electrochemical reduction of graphene oxide\cite{Carbon45;1558,RSCADV2;1168,ACSNano3;2653,AdvFunMat19;2782,
AdvFunMat20;3050}, and arc discharge\cite{STAM11;054502,AcsNano3;411,JPCC113;4257}.
The process of mechanical exfoliation is low-yield, size limited and cannot produce large scale graphenes. However, chemical methods have the advantage of yielding large-scale graphenes, but the drawback is the inferior quality of the exfoliated graphene sheets which contain many defects during the oxidation and reduction processes. Recently, chemical vapor deposition method promises as a potential means for large areas of high quality graphenes onto the transition-metal substrates (e.g., Ni, Pd, Pt, Ir, and Cu) based on the saturation of the hydrocarbon gas and the substrate\cite{JChemPhys129;234709,NanoLett9;30,ACSNano6;8241,ACSNano6;5680,
JPCC115;11666,Nature457;706,AdvMat26;5488,SurSci611;67}.
Previous works confirmed all the three highly symmetric stacking configurations, AA, AB and ABC, in the CVD-grown few-layer graphene\cite{PRL102;015501,ACSNano6;5680,PRB81;161410,JChemPhys129;234709,
JAP109;093523}.
Although AA configuration is theoretically predicted with higher total energy\cite{CARBON32;289}, the layer structure was grown on on the (111) surface of diamond\cite{JChemPhys129;234709} and on the SiC (000$\bar{1}$) substrate\cite{JAP109;093523}. Moreover this structure was frequently observed in bilayer graphenes\cite{PRL102;015501}. On the other hand, AB structure is mixed with ABC one\cite{NanoLett11;164,SurfSci610;53,Carbon46;272,PRB77;155426,NanoLett12;5539,
ACSNano6;5680} since the tiny difference between the total energies of the two configurations makes the difficulty of selectively producing  AB or ABC stacking\cite{PRB81;161410,CARBON32;289}.

AA-stacked graphenes have the simplest geometric structure, where the x- and y-coordinates of carbon atoms on different layers are vertically aligned\cite{CPL550;104,JChemPhys129;234709,JAP109;093523}.
Their band structure preserves the conical structure due to the vertically projected geometric symmetry; several Dirac cones are exactly located at the $K$ ($K^{'}$) point with energy separations in the order of the vertical interlayer hopping energy\cite{PRB83;165429,APL103;041907,JAP110;013725}.
In the AB stacking configuration, the second graphene sheet is shifted by one bond length along the armchair direction with respect to the first layer\cite{RevModPhys81;109}. If the third layer is further shifted in the same way, the arrangement sequence is known as an ABC
stacking\cite{RevModPhys81;109}.
AB- and ABC-stacked graphenes show more distinct band structures than AA ones as a result of the significant hybridization of energy bands under the interlayer atomic interactions\cite{PRB81;115315,PRB75;193402,PRB78;205425,PRL97;036803,Rep76;056503,
RevModPhys81;109,JPCM18;5849,PRL99;216802,JPSJ76;024701,
PRB84;165404,PRB82;035409,PRL73;245426,PRB83;220503,PRB87;155116}.
AB-stacked graphenes consist of several bilayer-like parabolic bands, while it owns a particular pair of monolayer-like linear bands if the total number of layers is odd\cite{Rep79;056503,PRB75;193402,PRB78;205425,PRB81;115315,RevModPhys81;109}.
In ABC systems, there is always one pair of weakly dispersive bands with surface-localized states near the Fermi level\cite{PRB84;165404,RevModPhys81;109,PRL73;245426,PRB83;220503,PRB87;155116}, and peculiar sombrero-shaped dispersion bands near the energy of the vertical nearest-neighboring interlayer interaction\cite{RevModPhys81;109,PRL97;036803,PRB84;165404,PRB82;035409}.
Several Dirac points are located at low energies for both AB- and ABC-stacked graphenes with their characteristics related to the number of layers and the stacking. The above-mentioned stacking-dependent electronic structures have been experimentally verified by infrared optical spectroscopy\cite{Nature4;532,PRB78;235408,PRB79;115441,PRL102;037403,PRL104;176404}. and angle-resolved photoemission spectroscopy\cite{PPL98;206802,PRB88;075406,PRB88;155439,PRL102;056808,Nature3;36,
PRB77;155303,PRL103;226803}

The electronic properties under the influence of external fields is focal point of some research\cite{PRL111;077402,PRL98;197403,Jnn11;4938,PRB84;075451,PRL102;256405,
PRB84;125455,JAP114;233701,CPL550;104,Rep76;056503,
JPCM18;5849,PRL99;216802,RevModPhys81;109,JPSJ76;024701,PRL97;266405,PhyE40;1722,
APL97;101905,JAP110;013725,PRL96;086805,PRB77;085426,PRB83;165443,PRB77;155416,
PRB84;205448,Ann326;721,RSCAdv4;56552,PRB80;165409,PRB90;205434,Carbon42;2975}.
In the presence of a uniform perpendicular electric field, the Dirac cones in AA-stacked graphenes are preserved\cite{CPL550;104,APL97;101905,JAP114;233701}, while only a rigid shift from $E_{F}=0$ takes place. However, the electric field can induce a band gap in few-layer AB- and ABC-stacked graphenes\cite{JPCM18;5849,JPSJ76;024701,PRL99;216802,PRB74;161403,PRB75;155115}.
Optical spectroscopy
\cite{PRL102;256405,PRB87;165102,Nature459;820,Science313;951} and
transport experiments\cite{NatNanotechnol4;383,Nature7;948,
SCIENCE320;206} confirmed that the gap can be continuously  tuned up to hundreds of meV. An tunable band gap makes them interesting for wide range of applications of electronic and photonic devices, such as electronic transistors\cite{NatCom3;906,NatTech2;605} and photodetectors\cite{NatTech9;780,NanoLett12;2773,NatPho4;611,NanoLett10;4285}.

Electronic states in a uniform perpendicular magnetic field $\mathbf{B}=B_{0}\hat{z}$ are evolved to quantized Landau levels (LLs)\cite{PRB77;155416,PRB77;085426,PRB80;165409,PRB83;165443,PRB84;161406,
PRB84;205448,PRB84;075451,PRB84;125455,PRB87;075417,PRB87;085424,PRB87;115422,
PRB90;205434,PRL96;086805,JAP110;013725,Ann326;721,RSCAdv4;56552,PhyE40;1722,
Jnn11;4938,Nature438;197,Carbon42;2975,APL97;101905}.
In density of states (DOS), the peculiar sequence of peaks is responsible for the quantization of massless Dirac carriers\cite{PRL109;176804,SCIENCE324;924,NATURE467;185,NATURE7;245,PRL109;116802}; this sequence is dependent on $\sqrt{nB_{0}}$, where $n$ is the quantum number and $B_{0}$ is the magnetic field strength\cite{PRL95;146801,Nature438;197,Nature438;201}.
On the other hand, the massive fermions in bilayer graphene give rise to quantized LL energies $\sqrt{n(n-1)}B_{0}$\cite{NATURE7;649,NatPhys2;177,PRL96;086805}.
The field-independent LL at $E_{F}=0$ for $n=0$ confirms that the Dirac points are hole-electron states in charge neutral graphene (available to either holes or electrons).
The LL spectra of multilayer graphenes are particularly depend on the layer number and stacking configuration\cite{PhyE40;1722,Carbon42;2975,APL97;101905,JAP110;013725,PRL96;086805,
PRB77;085426,PRB83;165443,PRB77;155416,PRB84;205448,PRB84;125455,Ann326;721,
Jnn11;4938,PRB80;165409,PRB90;205434,PRB84;161406,PRB87;075417,PRB84;075451,PRB87;085424,
PRB87;115422,RSCAdv4;56552}.
Based on the zero-field band structures, the LLs can be classified into several groups, with each being responsible for the magnetic quantization of the respective subband.
The nature of the Dirac points is the origin of the LLs with constant energy in monolayer graphene\cite{Science306;666,Proc102;10451,NatMat6;183,Nature438;197,Nature438;201} and with field-insensitive energies in few-layer graphenes\cite{APL97;101905,JAP110;013725,PRL96;086805,PRB77;085426,
PRB83;165443,PRB77;155416,PRB84;205448,PRB84;125455,Ann326;721,
Jnn11;4938,PRB80;165409,PRB90;205434}.
It should be noted that such LLs in AA-stacked graphenes are separated from each other by tens of meV\cite{JAP110;013725,CPL550;104}, while such LLs in AB- and ABC-stacked graphenes are confined in a narrow energy range of 10 meV around $E_{F}=0$\cite{PRL96;086805,PRB83;165443,Ann326;721,RSCAdv4;56552,PRB80;165409,PRB90;205434}; their energy differences can be verified by quantum transport experiments
\cite{PRB82;165404,PRB88;121302,NatPhys7;621,NatPhys7;953}.
Furthermore, one LL is coupled to another with a specific quantum number relationship, leading to the anticrossing patterns in LL spectra, which has been investigated in few-layer AB- and ABC-stacked graphenes and in the Bernal-stacked bulk graphite\cite{RSCAdv4;56552,PRB83;165443,PRB87;075417, PRB87;115422,JPSJ17;808}.
This special phenomenon can be explained by a perturbation of the non-vertical interlayer atomic interactions\cite{JPSJ17;808}. Previous works based on the comprehensive tight-binding model\cite{RSCAdv4;56552,PRB90;205434} showed that the Landau quantization of the sombrero-shaped subbands in ABC-stacked graphenes leads to the peculiar LL evolution under a variation of the magnetic field that reveals inversely field-dependent energies and a complex pattern of intragroup LL anticrossings, especially in the region of weak fields.

Scanning tunneling spectroscopy (STS) has been used to study the Dirac-fermion properties of graphenes through the measurements of the DOS\cite{PRL106;126802,PRB77;155426,SurfSci610;53,NanoLett3;5153,PRB76;201402,
NatMat12;887,NATURE6;109}.
In monolayer graphene, the DOS is linearly dependent on the energy and vanishes at the Dirac point, a behavior identical to that of the massless Dirac fermions\cite{PRL106;126802,NatMat12;887}.
It has been verified that massless and massive Dirac spectra can appear in AA- and AB-stacking bilayer graphenes, respectively\cite{NatMat12;887}. Furthermore, the coexistence of both spectra can be observed in symmetry-broken bilayer graphenes\cite{NATURE6;109,NatMat12;887}.
In a magnetic field, the LL energies can be directly measured through the energies of the sharp peaks in the differential conductance map of STS ($dI/dV$ versus $V$)\cite{SCIENCE324;924,NATURE467;185,NATURE7;245,PRL109;116802,PRL109;176804,
NATURE7;649}. The orbital Landau quantizations of massless Dirac fermions in monolayer\cite{SCIENCE324;924,NATURE467;185,NATURE7;245,PRL109;116802,PRL109;176804} and massive Dirac fermions in AB-stacked bilayer\cite{NATURE7;649} are, respectively, verified by the sequence of peaks with square-root and linear dependencies on $B_{0}$. In terms of the sequence of peaks, the indexes $n=0$ and $n=0, 1$ correspond to the monolayer and bilayer Dirac points, respectively.
On the other hand, the spectroscopic-imaging STM has been used to obtain the local DOS (LDOS) of LLs, such as in graphene\cite{NATURE6;811}, 2D electron gas\cite{PRL109;116805} and topological insulator\cite{NATURE10;815}, revealing the spatial distributions of the wave functions which consist of n-dependent nodal structures\cite{PRL109;116805,NATURE10;815}.
Owing to the LL anticrossings and the symmetry broken down in multilayer graphenes\cite{SCIENCE315;1379,PRB90;205434,PRB83;165443,PRB84;161406,
RSCAdv4;56552,PRB84;075451,PRB81;115315,PRB87;085424,PRB87;115422,
PRB87;075417}, it is interesting to measure the behavior of LLs and wave functions.

Optical spectroscopy can be utilized to identify the feature-rich electronic properties of graphenes\cite{Nature4;532,PRB78;235408,PRB79;115441,PRL102;037403,
PRL104;176404,PRL110;246803,PRL97;266405,PRL98;197403,PRL111;077402,PRL100;151,
PRL101;267601,PRB83;125302,PRB85;245410,PRL100;087403,PRL102;166401,PRB87;165102,
Nature459;820,PRL102;256405,Science313;951}.
The optical response is determined by vertical transitions from the occupied states to the unoccupied states.
Experimental\cite{PRL102;037403,PRB79;115441,Nature4;532,PRB78;235408,
PRL104;176404,PRB77;115313} and theoretical\cite{PRR89;045419,NJP15;015010,SynMet162;800,PRB83;245418,PRB73;144427} studies are consistent in the spectral absorption features of multilayer graphenes without external fields, such as the prominent structures, and the number, intensity and frequency of absorption peaks, which depend on the layer number and stacking configuration.
Furthermore, the theoretically predicted gap opening of biased AB-stacked bilayer and ABC-stacked trilayer graphenes has been verified through the observation of new transition channels within the energy gap\cite{PRB87;165102,
Nature459;820,PRL102;256405,Science313;951}.
The magnitude of the energy gap is compatible to that calculated within the tight-binding model over a wide spectral range\cite{JPCM18;5849,JPSJ76;024701,
PRB74;161403,PRB75;155115,PRB73;144427,PRL99;216802}.
The magneto-optical absorption frequencies, identified as monolayer and bilayer graphene features, show a square-root dependence\cite{PRL97;266405,PRL98;197403,PRL111;077402,PRL100;151,PRL101;267601}  and a linear dependence\cite{PRB73;245411,PRL98;157402,PRB83;125302,PRB85;245410,PRL100;087403,
ACSNano4;1465,PRB77;115313,PRL102;166401} on $B_{0}$, respectively.
The square-root dependent frequencies are also expected for the LL transitions of Dirac quasi-particles in the graphene-like 2D materials, e.g., MoS$_{2}$\cite{PRB89;155316,APLHo;acc} and silicene \cite{PRB88;085434,PRL110;197402} and the topological insulator \cite{PRB85;195440,PRB81;125120,PRB82;045122,APL100;161602}
According to the Fermi golden rule, the inter-LL transitions from the well-behaved LLs obey the specific selection rule $\Delta n=\pm1$ \cite{PRB89;155316,PRL110;197402,PRB82;045122,APL100;161602,APLHo;acc,PRL111;077402,PRL98;157402,
PRB73;245411,ACSNano4;1465,PRB77;115313}.
With an increase of the layer number, the magneto-electronic properties are enriched and diversified by distinct stackings.
A further exploration on the multilayer spectra needs to be undertaken in order to comprehend the magneto-optical response to the special LL spectra.

In electronic transport measurements, the peculiar Landau quantization of massless Dirac fermions results in equidistant plateaus of the Hall conductivity of $\sigma_{xy}=\pm4(n+1/2)e^{2}/h$, where $e^{2}/h$ is the conductance quantum and n is an integer number\cite{PRB82;165404,PRL95;146801,Nature438;201,NatMat6;183,Nature438;197,
Nanotechnology23;052001,SCIENCE315;1379}. The quantum Hall effect (QHE) can be observed even at room temperature due to the high mobility of electrons in graphene\cite{SCIENCE315;1379}. Unlike the situation in a conventional 2D electron gas (2DEG) system, the plateau at zero conductivity is missing due to the appearance of the $n=0$ LL at $E_{F}=0$\cite{Nature438;201,SCIENCE315;1379,Nanotechnology23;052001}.
While recent QHE experiments confirm the LL crossing, the plateau shift and the specific sequence of plateaus of the Hall conductivity in bilayer\cite{PRB82;165404,NatPhys2;177,NatMater7;151} and trilayer graphenes\cite{PRL107;126806,PRB88;121302,NatPhys7;621,NatPhys7;953,PRX2;011004,
NanoLett13;1627}, further explorations on the cases of distinct graphene systems are undertaken in order to comprehend the LL anticrossings and the unusually sequenced LLs.

The magneto-electronic properties are evidently derived within the framework of the generalized tight-binding model, which is applicable to multilayer systems with various stacking configurations and layer numbers. The sublattice dominance is determined by the subenvelop functions spanned by the bases in the tight-binding model.
Under a variation of the magnetic field strength, the LL spectra with respect to different stacking configurations can be realized. In terms of the relationship between the Landau state and the sublattice dominance, one can straightforwardly classify the LLs and define their quantum numbers. This provides a useful means for analyzing other essential properties of graphene systems. This review article broadly covers the fields related to multilayer graphene systems and discusses many factors affecting their magneto-electronic properties. In Chapter 2, we introduce the Peierls tight-binding model to obtain the Hamiltonian matrices describing the AA, AB and ABC stacking configurations, in the presence of a perpendicular uniform magnetic field. The Chapters 3-6 deal with the magneto-electronic properties of monolayer and multilayer graphenes with AA, AB and ABC stackings. We start with the case of monolayer graphene in Chapter 3, and subsequently discuss AA-stacked graphenes in Chapter 4, in which all layers are exactly aligned, leading to a symmetry-protected Dirac-like electronic spectrum. The LL spectra of both monolayer and AA-stacked multilayer graphenes show the square-root energy dependence on the field strength and the quantum number. Chapters 5 and 6 are devoted to AB- and ABC-stacked graphenes, which both show a non-monotonic LL energy dependence, and reveal a complex pattern of LL anticrossings as a result of certain interlayer atomic interactions. In each of these chapters, we present the theoretical calculations of the LL spectra and the wave functions, certain important characteristics of which have been confirmed experimentally by STS, optical spectroscopy and quantum transport measurements (described in the final subchapter). We focus on magneto-absorbtion experiments, which are available for the investigation of the layer and stacking dependence of the magneto-electronic properties over a broad range of energies. Meanwhile, a detail comparison among these graphene systems is also presented in Chapter 7. Finally, Chapter 8 contains concluding remarks.

\bigskip
\bigskip
\section{The generalized tight-binding model}
\bigskip
\bigskip

Electronic states are evolved to discrete dispersionless LLs in the presence of a magnetic field. In this chapter, we show how to calculate the Landau energy bands of multilayer graphenes in the framework of the generalized tight-binding model, which is based on the subenvelope functions of different sublattices. The Hamiltonian is built from the bases of tight-binding functions coupled with a Pierer phase factor. Three typical kinds of stacking configurations are used as study models: AA, AB and ABC. The magnetic quantization of electrons in graphene systems show interesting phenomena as a function the stacking configuration and the number of graphene layers.

\subsection{Monolayer graphene}

An illustration of the honeycomb structure of monolayer graphene is shown in Fig. 1(a).
The primitive unit cell containing two carbon atoms, $A$ (black) and $B$ (red) atoms, is marked by the gray shadow. The symbols $\alpha _{0}$ ($\simeq-2.6$ eV) and $b^{'}$ ($=1.42$ \AA ) indicate the nearest-neighbor hopping integral and C-C bond length, respectively.
Furthermore, $\alpha _{0}$ is the most important atomic interaction often introduced to calculate the physical properties of monolayer graphene.
The Brillouin zone corresponding to the primitive unit cell is shown in Fig. 1(b), where
$\mathbf{\Gamma }$, $\mathbf{M}$ and $\mathbf{K}$ are three highly symmetric points. Based on the primitive unit cell shown in Fig. 1(a), the wave function is expressed as $\Psi =\varphi _{A}+\lambda \varphi_{B}$, where $\varphi _{A}$ and $\varphi _{B}$, respectively, stand for the Bloch wave functions of the $A$ and $B$ atoms and are represented as\cite{PR71;622}
\begin{eqnarray}
\begin{array}{l}
\varphi _{A} =A\exp (i\mathbf{k}\cdot \mathbf{R}_{A})\chi (%
\mathbf{r}-\mathbf{R}_{A})  \label{1a} \\
\varphi _{B} =B\exp (i\mathbf{k}\cdot \mathbf{R}_{B})\chi (%
\mathbf{r}-\mathbf{R}_{B})\text{.}  \label{1b}
\end{array}
\end{eqnarray}%
$\chi (\mathbf{r}-\mathbf{R}_{A})$ ($\chi (\mathbf{r}-\mathbf{R}_{B})$) is
the normalized orbital 2$p_{z}$ wave function for an isolated atom at $%
\mathbf{R}_{A}$\ ($\mathbf{R}_{B}$), with $\lambda =\pm 1$ indicating the
bonding and anti-bonding forms.

In the presence of a uniform perpendicular magnetic field $\mathbf{B}=B_{0}\widehat{z}$,
the vector potential is chosen as $\mathbf{A}(\mathbf{r})\mathbf{=}B_{0}x%
\widehat{y}$\ and the related periodic Peierls phase$\Delta
G_{mm^{\prime }}\equiv {\frac{2{\pi }}{{\phi }_{0}}}$ $\int_{\mathbf{R}%
_{m^{\prime }}}^{\mathbf{R}_{m}}\mathbf{A}(\mathbf{r})\cdot d\mathbf{r}$ \
is introduced in the tight-binding functions, where $\phi _{0}=hc/e$ ($4.1356\times
10^{-15}$ $[$T m$^{2}]$) is the flux quantum\cite{Ann326;721,Jnn11;4938,PRB90;205434}.
Owing to the periodicity of the Peierls phase, the primitive unit cell becomes an enlarged rectangle unit cell along the $x$-direction (armchair direction), as indicated in Fig. 1(c), where $R_{B}$ is associated with the period along $\hat{x}$ and defined as $R_{B}=\frac{\phi _{0}/(3\sqrt{3}b^{\prime 2}/2)}{B_{0}}\simeq \frac{79000\text{ T}}{B_{0}}$.
Accordingly, the rectangle cell includes $4R_{B}$ atoms$\ $($2R_{B}$ $A$ and $2R_{B}$ $B$ atoms); its length along the $x$-direction is given by $l=3R_{B}b$.
This implies that the Bloch wave functions under a uniform magnetic field can be expressed by the linear superposition of the $4R_{B}$ Peierls tight-binding functions in the rectangular unit cell: $|A_{1\mathbf{k}}\rangle $, $|B_{1\mathbf{k}}\rangle $, $|A_{2\mathbf{k}}\rangle $, $|B_{2\mathbf{k}}\rangle $, ......$|A_{2R-1\mathbf{k}}\rangle $, $|B_{2R-1\mathbf{k}}\rangle $, $|A_{2R\mathbf{k}}\rangle $; $|B_{2R\mathbf{k}}\rangle $.
To solve the $4R_{B}\times 4R_{B}$\ Hamiltonian matrix more efficiently, a band-like Hamiltonian matrix\cite{PRB77;085426,Ann326;721,Jnn11;4938,PRB90;205434} is introduced by rearranging the bases as the sequence $|A_{1\mathbf{%
k}}\rangle $, $|B_{2R\mathbf{k}}\rangle $, $|B_{1\mathbf{k}}\rangle $, $%
|A_{2R\mathbf{k}}\rangle $, $|A_{2\mathbf{k}}\rangle $, $|B_{2R-1\mathbf{k}%
}\rangle $, $|B_{2\mathbf{k}}\rangle $, $|A_{2R-1\mathbf{k}}\rangle $, ......%
$|A_{R-1\mathbf{k}}\rangle $, $|B_{R+2\mathbf{k}}\rangle $, $|B_{R-1\mathbf{k%
}}\rangle $, $|A_{R+2\mathbf{k}}\rangle $, $|A_{R\mathbf{k}}\rangle $, $%
|B_{R+1\mathbf{k}}\rangle $, $|B_{R\mathbf{k}}\rangle $; $|A_{R+1\mathbf{k}%
}\rangle $. This matrix is thus expressed as

\begin{equation}
\left[
\begin{array}{cccccccc}
0 & q^{\ast } & p_{1}^{\ast } & 0 & \cdots & 0 & 0 & 0 \\
q & 0 & 0 & p_{2R} & 0 & \cdots & 0 & 0 \\
p_{1} & 0 & 0 & 0 & \ddots & \ddots & \vdots & 0 \\
0 & p_{2R}^{\ast } & 0 & 0 & \ddots & \ddots & 0 & \vdots \\
\vdots & 0 & \ddots & \ddots & \ddots & 0 & p_{R}^{\ast } & 0 \\
0 & \vdots & \ddots & \ddots & 0 & 0 & 0 & p_{R+1} \\
0 & 0 & \cdots & 0 & p_{R} & 0 & 0 & q \\
0 & 0 & 0 & \cdots & 0 & p_{R+1}^{\ast } & q^{\ast } & 0%
\end{array}%
\right] \text{.}
\end{equation}%
The elements $p_{m}$ and $q$ are defined as
\begin{eqnarray}
\begin{array}{l}
p_{m} \equiv \gamma _{0}\exp -i
\left[ \left( k_{x}b/2+k_{y}\sqrt{3}b/2+\Delta G_{mm^{\prime }}\right)
+\left( k_{x}b/2-k_{y}\sqrt{3}b/2-\Delta G_{mm^{\prime }}\right) \right] ,  \\
q \equiv \gamma _{0}\exp [i(k_{x}b)]\text{;}   \\
\Delta G_{mm^{\prime }} =-\pi (m-5/6)/R\text{.}
\end{array}
\end{eqnarray}
By diagonalizing the Hamiltonian matrix, the energy dispersion $E^{c,v}$ and the wave function $\Psi ^{c,v}$ are obtained.

\subsection{Multilayer graphenes}

When graphene layers are stacked along the $z$-direction, the physical properties of an $N$-layer graphene are strongly affected by its layer number, interlayer atomic interactions and stacking configurations\cite{NJP15;015010,SynMet162;800,PRB83;245418,PRB73;144427,CARBON69;151,
PRB73;245411,PRL98;157402,ACSNano4;1465,PRB77;115313,PRB86;125434,ACSnano5;1026,NanoResLett7;134,PRB89;165407,
PRL95;146801,Nature438;201,SCIENCE315;1379,Nanotechnology23;052001,
PRB82;165404,NatPhys2;177,NatMater7;151,NatPhys7;621,PRL107;126806,PRB88;121302,
NatPhys7;953,NanoLett13;1627,PRX2;011004}. Three typical stacking configurations, namely AA, AB and ABC, are usually chosen for theoretical and experimental studies. They present distinct physical properties, e.g., optical absorption spectra\cite{NJP15;015010,SynMet162;800,PRB83;245418,PRB73;144427,CARBON69;151,
PRB73;245411,PRL98;157402,ACSNano4;1465,PRB77;115313}, Coulomb excitations\cite{PRB86;125434,ACSnano5;1026,NanoResLett7;134,PRB89;165407} and the quantum Hall effects\cite{PRL95;146801,Nature438;201,SCIENCE315;1379,Nanotechnology23;052001,
PRB82;165404,NatPhys2;177,NatMater7;151,NatPhys7;621,PRL107;126806,PRB88;121302,
NatPhys7;953,NanoLett13;1627,PRX2;011004}, which do reflect the influences of layer number, stacking configurations and interlayer interactions. In this chapter, the Hamiltonian matrix of the AA-, AB- and ABC-stacked $N$-layer graphenes are derived in the following.

\subsubsection{AA stacking}

The geometric structure of an AA-stacked $N$-layer graphene is illustrated on the top right side in Fig. 2.
With an AA-stacking configuration, the geometric structure is formed by periodically stacked monolayer graphenes along the $\hat{z}$ direction.
Four important atomic interactions are included in the tight-binding calculations: one nearest-neighbor intralayer interaction $\alpha _{0}=2.569$ eV, and three interlayer interactions $\alpha _{1}=0.361$eV, $\alpha _{2}=0.013$ eV and $\alpha _{3}=-0.032$ eV\cite{PRB46;4531}. The primitive unit cell under a uniform magnetic field consists of $N\times 4R_{B}$ atoms. In order to obtain a band-like matrix, we arrange the bases as follows: $|A_{1\mathbf{k}}^{1}\rangle $, $|B_{1\mathbf{k}}^{1}\rangle $, $|A_{1\mathbf{k}}^{2}\rangle $, $|B_{1\mathbf{k}}^{2}\rangle $,...$|A_{1\mathbf{k}}^{N-1}\rangle $, $|B_{1\mathbf{k}}^{N-1}\rangle $, $|A_{1\mathbf{k%
}}^{N}\rangle $, $|B_{1\mathbf{k}}^{N}\rangle $,.....$|A_{2\mathbf{k}%
}^{1}\rangle $, $|B_{2\mathbf{k}}^{1}\rangle $, $|A_{2\mathbf{k}}^{2}\rangle
$, $|B_{2\mathbf{k}}^{2}\rangle $,...$|A_{2\mathbf{k}}^{N-1}\rangle $, $|B_{2%
\mathbf{k}}^{N-1}\rangle $, $|A_{2\mathbf{k}}^{N}\rangle $, $|B_{2\mathbf{k}%
}^{N}\rangle $,.....$|A_{2R_{B}\mathbf{k}}^{1}\rangle $, $|B_{2R_{B}\mathbf{k}%
}^{1}\rangle $, $|A_{2R_{B}\mathbf{k}}^{2}\rangle $, $|B_{2R_{B}\mathbf{k}%
}^{2}\rangle $,...$|A_{2R_{B}\mathbf{k}}^{N-1}\rangle $, $|B_{2R_{B}\mathbf{k}%
}^{N-1}\rangle $, $|A_{2R_{B}\mathbf{k}}^{N}\rangle $; $|B_{2R_{B}\mathbf{k}%
}^{N}\rangle $. One thing that should be noted is that all hopping terms and external fields considered in calculations regarding $N$-layer graphenes are not just treated as perturbation terms; rather, the effects of atomic interactions and external fields can be simultaneously included in the calculations. For an AA-stacked $N$-layer graphene, the nonzero matrix elements are
\begin{eqnarray}
\begin{array}{l}
\langle B_{m\mathbf{k}}^{L}|H|A_{m^{\prime }\mathbf{k}}^{L^{\prime }}\rangle
=\alpha _{0}(t_{1k}\delta _{m^{\prime },m}+t_{4k}\delta _{m^{\prime
},m+1})\delta _{L,L^{\prime }}+\alpha _{3}(t_{1k}\delta _{m^{\prime
},m}+t_{4k}\delta _{m^{\prime },m-1})\delta _{L-1,L^{\prime }}\\
+\alpha_{3}(t_{1k}\delta _{m^{\prime },m}+t_{4k}\delta _{m^{\prime },m+1})\delta
_{L+1,L^{\prime }}\text{,}  \label{4.1} \\
\langle A_{m\mathbf{k}}^{L}|H|A_{m^{\prime }\mathbf{k}}^{L^{\prime }}\rangle
=\langle B_{m\mathbf{k}}^{L}|H|B_{m^{\prime }\mathbf{k}}^{L^{\prime
}}\rangle =\alpha _{1}\delta _{L,L^{\prime }+1}+\alpha _{2}\delta
_{L,L^{\prime }+2}\text{,}  \label{4.2} \\
\end{array}
\end{eqnarray}
where the superscript in the tight-binding function represents the Lth layer. The four independent phase terms are
\begin{eqnarray}
\begin{array}{l}
t_{1k} ={\exp }\{i[-(k_{x}b/2)-(\sqrt{3}k_{y}b/2)]+\pi(m-5/6)/R\}\\
+{\exp }\{i[-(k_{x}b/2)+(\sqrt{3}k_{y}b/2)]-\pi (m-5/6)/R\}\text{,} \label{4.5} \\
t_{2k} ={\exp }\{i[-(k_{x}b/2)-(\sqrt{3}k_{y}b/2)]+\pi (m-3/6)/R\}\\
+{\exp }\{i[-(k_{x}b/2)+(\sqrt{3}k_{y}b/2)]-\pi (m-3/6)/R\}\text{,} \label{4.6} \\
t_{3k} ={\exp }\{i[-(k_{x}b/2)-(\sqrt{3}k_{y}b/2)]+\pi (m-1/6)/R\}\\
+{\exp }\{i[-(k_{x}b/2)+(\sqrt{3}k_{y}b/2)]-\pi (m-1/6)/R\}\text{,} \label{4.7} \\
t_{4k} ={\exp }(ik_{x}b)\text{.}  \label{4.8}
\end{array}
\end{eqnarray}

When an electric field is applied, the electric potentials are added to the site energies of the diagonal Hamiltonian matrix elements.

\subsubsection{AB stacking}

The diagrammatic drawing of an $N$-layer graphene in the AB-stacking configuration is shown in Fig. 3.
Each of the graphene sheets is obtained by shifting adjacent layers along the armchair direction by one bond length $b^{'}$. The $A$ atoms on any odd-number layer have the same $(x,y)$ coordinates as $B$ atoms on the upper layer, while its $B$ atoms are projected at the center of the hexagon of the adjacent layer. The atomic interactions based on the SWMcC model include one nearest-neighbor intralayer interaction $\gamma _{0}=-3.12$ eV, five interlayer interactions, $\gamma _{1}=0.38$ eV, $\gamma _{2}=-0.021$eV, $\gamma _{3}=0.28$ eV, $\gamma _{4}=0.12$ eV, $\gamma _{5}=-0.003$ eV, and the chemical environment difference between $A$ and $B$ $\gamma _{6}=-0.0366$ eV\cite{PRB46;4540,CARBON32;289}. The primitive unit cell of AB-stacked $N$-layer graphene consists of $2\times N$\ atoms, and the magnetically enlarged rectangular cell comprises $N\times 4R_{B}$ atoms. Unlike the effective-mass calculation that treat non-vertical interactions as perturbation terms, such as $\gamma _{3}$ and $\gamma _{4}$, the tight-binding calculations can well define the dimensions of the Hamiltonian matrix by simultaneously considering all the all SWMcC parameters and the magnetic field. The Hamiltonian matrix elements are expressed as
\begin{eqnarray}
\begin{array}{l}
\langle A_{m}^{L}|H|A_{m^{\prime }}^{L^{\prime }}\rangle =\gamma
_{1}\delta _{m,m^{\prime }}\delta _{L,L^{\prime }\pm 1}+(\gamma _{5}+\gamma
_{6})\delta _{m,m^{\prime }}\delta _{L,L^{\prime }}+{\frac{\gamma _{5}}{2}}%
\delta _{m,m^{\prime }}\delta _{L,L^{\prime }\pm 2},  \label{6.1} \\
\langle B_{m}^{L}|H|B_{m^{\prime }}^{L^{\prime }}\rangle =\gamma
_{3}(t_{2,k}\delta _{m,m^{\prime }}+t_{4,k}^{\ast }\delta _{m,m^{\prime
}-1})\delta _{L,L^{\prime }\pm 1}+\gamma _{2}\delta _{m,m^{\prime }}\delta
_{L,L^{\prime }}+{\frac{\gamma _{2}}{2}}\delta _{m,m^{\prime }}\delta
_{L,L^{\prime }\pm 2},  \label{6.2} \\
\langle A_{m}^{L}|H|B_{m^{\prime }}^{L^{\prime }}\rangle
=\gamma_{0}(t_{4,k}\delta _{m,m^{\prime }}+t_{3,k}^{\ast }\delta _{m,m^{\prime
}+1})\delta _{L,L^{\prime }}+\gamma _{4}(t_{1,k}\delta _{m,m^{\prime
}}+t_{4,k}^{\ast }\delta _{m,m^{\prime }+1})\delta _{L,L^{\prime }\pm 1}%
\text{ for odd }L,  \label{6.3} \\
=\gamma _{0}(t_{1,k}\delta _{m,m^{\prime }}+t_{4,k}^{\ast }\delta
_{m,m^{\prime }+1})\delta _{L,L^{\prime }}+\gamma _{4}(t_{4,k}\delta
_{m,m^{\prime }}+t_{3,k}^{\ast }\delta _{m,m^{\prime }+1})\delta
_{L,L^{\prime }\pm 1}\text{ for even L.}  \label{6.4}
\end{array}
\end{eqnarray}

\subsubsection{ABC stacking}

The ABC stacking configuration is illustrated in Fig. 4, where each of the graphene sheets is shifted by a distance of $b_{'}$ along the armchair direction with respect to the adjacent layer.
The sublattice $B$ of one layer is situated directly above the $A$ atom of the adjacent lower layer, whereas sublattice $A$ lies above the center of its hexagon. The calculation for such a system takes one nearest-neighbor intralayer interaction $\beta _{0}=-3.16$ eV and five interlayer interactions $\beta_{1}=0.36$ eV, $\beta _{2}=-0.01$ eV, $\beta _{3}=0.32$ eV, $\beta_{4}=0.03$ eV and $\beta _{5}=0.0065$ eV, into account\cite{PRB46;4540}. Consequently, the magnetically enlarged rectangular cell of an ABC-stacked $N$-layer graphene is composed of $N\times 4R_{B}$ atoms and the Hamiltonian matrix, built by the $4NR_{B}$ bases, is represented as

\begin{eqnarray}
\begin{array}{l}
\langle B_{m\mathbf{k}}^{L}|H|A_{m^{\prime }\mathbf{k}}^{L^{\prime }}\rangle\\
=\beta _{0}(t_{1,k}\delta _{m,m^{\prime }}+t_{4,k}\delta _{m,m^{\prime
}-1})\delta _{L,L^{\prime }}+\beta _{3}(t_{3,k}^{\ast }\delta _{m,m^{\prime
}+1}+t_{4,k}^{\ast }\delta _{m,m^{\prime }})\delta _{L,L^{\prime }-1}\text{
for $L=3l-2$}, \\
\langle B_{m\mathbf{k}}^{L}|H|A_{m^{\prime }\mathbf{k}}^{L^{\prime }}\rangle\\
=\beta _{0}(t_{3,k}\delta _{m,m^{\prime }+1}+t_{4,k}\delta _{m,m^{\prime
}})\delta _{L,L^{\prime }}+\beta _{3}(t_{2,k}^{\ast }\delta _{m,m^{\prime
}}+t_{4,k}^{\ast }\delta _{m,m^{\prime }-1})\delta _{L,L^{\prime }-1}\text{
for }L\text{$=3l-1$}, \\
\langle B_{m\mathbf{k}}^{L}|H|A_{m^{\prime }\mathbf{k}}^{L^{\prime }}\rangle\\
=\beta _{0}(t_{2,k}\delta _{m,m^{\prime }}+t_{4,k}\delta _{m,m^{\prime
}-1})\delta _{L,L^{\prime }}+\beta _{3}(t_{1,k}^{\ast }\delta _{m,m^{\prime
}}+t_{4,k}^{\ast }\delta _{m,m^{\prime }-1})\delta _{L,L^{\prime }-1}\text{
for }L\text{$=3l$}, \\
\langle B_{m\mathbf{k}}^{L}|H|A_{m^{\prime }\mathbf{k}}^{L^{\prime }}\rangle\\
=\beta _{2}\delta _{m,m^{\prime }}\delta _{L,L^{\prime }-2}+\beta
_{1}\delta _{m,m^{\prime }}\delta _{L,L^{\prime }-1}\text{ for any L},\\
\langle B_{m\mathbf{k}}^{L}|H|B_{m^{\prime }\mathbf{k}}^{L^{\prime }}\rangle
=\langle A_{m\mathbf{k}}^{L+2}|H|A_{m^{\prime }\mathbf{k}}^{L^{\prime}+2}\rangle\\
=\beta _{4}(t_{1,k}\delta _{m,m^{\prime }}+t_{4,k}\delta
_{m,m^{\prime }-1})\delta _{L,L^{\prime }-1}\text{ for $L=3l-2$}, \\
\langle A_{m\mathbf{k}}^{L}|H|A_{m^{\prime }\mathbf{k}}^{L^{\prime }}\rangle
=\langle B_{m\mathbf{k}}^{L+1}|H|B_{m^{\prime }\mathbf{k}}^{L^{\prime}+1}\rangle\\
=\beta _{4}(t_{3,k}\delta _{m,m^{\prime }+1}+t_{4,k}\delta
_{m,m^{\prime }})\delta _{L,L^{\prime }-1}\text{ for $L=3l-2$}, \\
\langle A_{m\mathbf{k}}^{L}|H|B_{m^{\prime }\mathbf{k}}^{L^{\prime }}\rangle
=\langle B_{m\mathbf{k}}^{L^{\prime }+2}|H|B_{m^{\prime }\mathbf{k}%
}^{L+2}\rangle=\langle A_{m\mathbf{k}}^{L^{\prime }+1}|H|A_{m^{\prime }%
\mathbf{k}}^{L+1}\rangle\\
=\beta _{5}(t_{3,k}\delta _{m,m^{\prime
}+1}+t_{4,k}\delta _{m,m^{\prime }})\delta _{L,L^{\prime }-2}
\text{ for $L=3l-2,$} \\
\langle A_{m\mathbf{k}}^{L}|H|A_{m^{\prime }\mathbf{k}}^{L^{\prime }}\rangle
=\langle B_{m\mathbf{k}}^{L+1}|H|B_{m^{\prime }\mathbf{k}}^{L^{\prime}+1}\rangle\\
=\beta _{4}(t_{2,k}\delta _{m,m^{\prime }}+t_{4,k}\delta
_{m,m^{\prime }-1})\delta _{L,L^{\prime }-1}\text{ for $L=3l-1$}, \\
\langle B_{m\mathbf{k}}^{L}|H|B_{m^{\prime }\mathbf{k}}^{L^{\prime }}\rangle
=\langle A_{m\mathbf{k}}^{L+2}|H|A_{m^{\prime }\mathbf{k}}^{L^{\prime
}+2}\rangle =\langle A_{m\mathbf{k}}^{L^{\prime }+1}|H|B_{m\mathbf{k}%
}^{L+1}\rangle \\
=\beta _{5}(t_{2,k}\delta _{m,m^{\prime }}+t_{4,k}\delta
_{m,m^{\prime }-1})\delta _{L,L^{\prime }+2}\text{ for $L=3l$}, \\
\langle A_{m\mathbf{k}}^{L}|H|A_{m^{\prime }\mathbf{k}}^{L^{\prime }}\rangle
=\langle B_{m\mathbf{k}}^{L+1}|H|B_{m^{\prime }\mathbf{k}}^{L^{\prime
}+1}\rangle=\langle A_{m\mathbf{k}}^{L^{\prime }+2}|H|B_{m^{\prime }\mathbf{%
k}}^{L+2}\rangle \\
=\beta _{5}(t_{1,k}\delta _{m,m^{\prime }}+t_{4,k}\delta
_{m,m^{\prime }-1})\delta _{L,L^{\prime }+2}\text{ for $L=3l;$} \\
\end{array}
\end{eqnarray}

The generalized tight-binding model can deal with the effects of several kinds of external fields in the calculations, e.g. modulated magnetic fields\cite{PRB83;195405}, periodic potentials\cite{JVSTB28;386} and composite fields\cite{OptEx22;7473}. An exact diagonalization method that transforms the Hamiltonian matrix into a band-like one is utilized to characterize the magnetic quantum numbers and wave functions. The effects of external fields and geometric structures are directly reflected in the electronic properties. The physical phenomena are explain in a clear physical picture.
This model can be further developed to investigate magneto-optical properties and Coulomb excitations, which both are closely related to Landau-level spectra and transition matrix elements. The transition intensity needs to be evaluated from the initial- and final-state wave functions, in which each individual wave function is a linear combination of the products between the subenvelop function and the tight-binding function on each site in the magnetically enlarged unit cell.

Generally, the generalized tight-binding model with an effective digitalization method is applicable to study the quantization effect in arbitrarily stacked graphenes, other layered materials, e.g. MoS$_{2}$\cite{PRL105;136805,NatTech6;147,PRB89;155316,APLHo;acc} and silicene\cite{APL96;183102,PRL108;155501,PRL108;245501}, and bulk topological insulator\cite{PRL109;166407,NATURE5;438,NATURE452;970}. The characteristics of electronic structures and wave functions are well depicted. Moreover, the results are accurate and reliable within a wide energy range. On the other hand, the effective-mass model expands the low-energy Hamiltonian near the K point, deriving an analogue of the Dirac Hamiltonian for massless fermions, where the interlayer atomic interactions and external fields are regarded as perturbations in the calculations. It could conceivably be used to comprehend the magnetic quantization at low energies for fewer-layer graphenes with symmetric structures.

\bigskip
\bigskip
\section{Magnetic quantization in monolayer graphene}
\subsection{Dirac-cone band structure}
\bigskip
\bigskip

Monolayer graphene exhibits the Dirac-type linear dispersion near $E_{F}=0$ due to the special hexagonal lattice structure, as shown in Fig. 5.
A magnetic field causes the states to congregate and induces dispersionless LLs, as shown in Fig. 6(a) for $B_{0}=20$ T.
The unoccupied LLs and occupied LLs are symmetric about $E_{F}=0$, and characterized by quantum numbers $n^{c,v}$, which corresponds to the number of zeros in the wave function. Each Dirac-LL is fourfold degenerate without considering the spin degeneracy. Its energy can be described by the simple square-root relationship $|E_{n}^{c,v}|\propto \sqrt{n^{c,v}B_{0}}$.

\subsection{Landau levels and Landau wave functions}
For understanding the magneto-electronic properties, researching the evolution of the LL wave functions is very useful. Based on the arrangement of odd-indexed and even-indexed atoms in the primitive cell (Fig. 1(c)), the wave function can be expressed as%
\begin{equation}
|\Psi _{\mathbf{k}}\rangle =\sum\limits_{m=1}^{2R_{B}-1}(A_{o}^{c,v}|a_{m\mathbf{k}}\rangle +B_{o}^{c,v}|b_{m\mathbf{k}}\rangle
)+\sum\limits_{m=1}^{2R_{B}}(A_{e}^{c,v}|a_{m\mathbf{k}}\rangle +B_{e}^{c,v}|b_{m\mathbf{k}}\rangle )\text{,}
\end{equation}%
where $o (e)$ represents an odd (even) integer. Subenvelop functions of $A_{\mathbf{o}}^{c,v}$ ($A_{\mathbf{e}}^{c,v}$) and $B_{\mathbf{o}}^{c,v}$ ($B_{\mathbf{e}}^{c,v}$)
stand for the probability amplitudes of the wave functions of the $A$ and $B$ atoms, respectively. The wave functions associated with the odd-indexed and even-indexed atoms in Eq. (7) have only a phase difference of $\pi $, that is, $A_{o}^{c,v}=-A_{e}^{c,v}$ and $B_{o}^{c,v}=-B_{e}^{c,v}$. Therefore, it is sufficient to comprehend the main characteristics of the LLs by discussing only the amplitudes of $A_{o}^{c,v}$ and $B_{o}^{c,v}$. As shown in Figs. 6(b) and 6(c), the subenvelop functions exhibit even and odd spatial symmetries and can be described by an $n^{c,v}$-th order Hermite polynomial and Gaussian function.
They are distributed around the localization center, that is, at the $5/6$ position of the enlarged unit cell. On the other hand, similar localization centers corresponding to the other degenerate states occur at the $1/6$, $2/6$, and $4/6$ positions. A simple relationship of subenvelop functions exists between the $A$ and $B$ sublattices, i.e., $A_{o}^{c,v}$ of $n^{c,v}$ is linearly proportional to $B_{o}^{c,v}$ of $n^{c,v}+1$. Moreover, the conduction and valence wave functions are related to each other by the following relationship: $A_{o}^{c}=A_{o}^{v}$ and $B_{o}^{c}=-B_{o}^{v}$.

The dependences of the LL energies on $n^{c,v}$ and $B_{0}$ behave differently at low and high energies, as shown in Figs. 7(a) and 7(b).
The LLs with $E_{n}^{c}\lesssim 0.4\gamma _{0}$ are proportional to the product of $\sqrt{n^{c}}$ and to $\sqrt{B_{0}}$ simultaneously, but there is no such relation for the LLs with $E_{n}^{c}\gtrsim 0.4\gamma _{0}$, because the high-energy LLs are too densely packed to be separated from one another\cite{PhyE40;1722}.

\subsection{Oscillating Landau subbands under a modulated magnetic field}
As compared with the uniform case, a modulated magnetic field has a different impact on the electronic properties. The spatially periodic field along the armchair direction is assumed to be $B_{M}\sin (2\pi x/3b^{\prime }R_{M}) \widehat{z}$, resulting in oscillating Landau subbands, as shown in Fig. 8 for the modulation field strength $B_{M}=20$ T and the period $R_{M}=500$.
In the lower energy region, the conduction and valence subbands, symmetric about $E_{F}=0$, appear around $k_{y}=2/3$.
The subbands at $E_{F}=0$ are partially flat.
For the other Landau subbands with weak energy dispersions, they are characterized by double degeneracy and have one band-edge state at $k_{y}=2/3$ and four extra band-edge states on both sides of $k_{y}=2/3$.
The former band-edge state energies at $k_{y}=2/3$ are close to the LL energies, while the latter state energies are dependent on the modulation field strength and period.
They demonstrate the strongest dispersion and the destruction of the double degeneracy. Moreover, the $k_{y}$ range with respect to both the weak dispersions and partial flat
bands grows with an increasing field strength and a longer modulation period. On the contrary, when the influence of the modulation field becomes much weaker with increasing energy, the parabolic subbands in the higher energy region are similar to the twofold degenerate subbands directly obtained by the zone folding of monolayer graphene in the $B_{M}=0$ case.

\subsection{Geometry-induced non-uniform magnetic field in curved graphene systems}

Investigations on curved systems (e.g., curved graphene nanoribbons and carbon nanotubes\cite{CARBON69;151,JPSJ65;505,PRB67;045405,PRB62;16092,Science304;1129}) provide another means of studying the effect of a non-uniform magnetic field on graphene systems, and the cooperation and competition between the geometry and a magnetic field. In finite-size 1D graphene nanoribbons, the quantum confinement limits the formation of Landau states in the low-energy region\cite{PRB59;8271,Nanotechnology18;495401,
PRL107;086601,PRL110;246803}, where the energy bands are a combination of quaisi-LLs and parabolic bands, as shown in Fig. 9(a).
The qausi-LLs possess the dispersionless energy parts in a constrained wave-vector range, whereas it increases in the cases of stronger magnetic flux, i.e., stronger fields or wider ribbons. As a flat ribbon is bent into a curved one, the quasi-LLs are depressed due to a weak magnetic quantization. As shown in Fig. 9(b), the Quasi-LLs entirely evolve into oscillating Landau subbands, of which the energy dispersion and the band-edge states are determined by the confinements of both the open edge and the local magnetic field. On the other hand, in a carbon nanotube, a cylindrical structure corresponding to a zipped curved graphene ribbon, the zero net magnetic flux enhances the difficulty in forming the quasi-LLs. Instead, a transverse magnetic field results in a coupling of angular-momentum states, as shown in Fig. 9(c). With the magnetic field varying from low- to high- strength regimes, the coupling of independent angular momenta becomes significant. Recently in an extremely severe condition, i.e., the nanotube diameter is larger than the magnetic length, an experimental evidence of the Landau states in multi-walled carbon nanotubes under a very strong field, as high as 60 T, was identified by magneto-transport measurements\cite{PRL101;046803,PRL103;256801}.

Graphene, a flexible sheet of carbon atoms, can be curled or twisted without loosing its atomic structure and unique electronic properties. To date, a series of curled and folded graphene-related materials that posses specific geometric symmetries and dimensions have been synthesized---such as graphene bubbles\cite{NatCom4;1556}, graphene ripples\cite{NatTech4;562,NJP12;093018}, carbon tori\cite{Nat385;780}, carbon nanoscrolls\cite{PRL108;166602}, coiled carbon nanotubes\cite{Carbon48;3931,NJP13;063047,PRL105;106802,PRB83;165403,PRB84;041404}. Of further research interest are firstly the quantization of electronic states resulting from the cooperation and competition between the geometry and a magnetic field, and secondly whether the quasi-Landau states can survive or not.

\section{ Magnetic quantization in AA-stacked graphene}

AA-stacked graphene possesses a highly symmetric geometry structure, in which the carbon atoms of one layer are directly above/below the carbon atoms of another layer.
In natural graphite, graphene layers are normally stacked in AB and ABC orders, while AA stacking configuration is not found due to the much higher total energy in comparison with AB and ABC configurations. Recently, however, it can be artificially synthesized and maintain the stability for macroscopically long times\cite{JChemPhys129;234709,JAP109;093523}. The electronic band structure in the low-energy region comprises several Dirac cones\cite{PRB83;165429}. The massless Dirac fermions with linear dispersion were verified by STS experiments on the bilayer AA-stacked graphene\cite{NatMat12;887}. The nature of the Dirac fermions attracts intense interest of theoretical and experimental research in AA-stacked graphenes. In this chapter, we mainly focus on how the low-lying electronic states within the Dirac cone are quantized by a uniform perpendicular magnetic field. Furthermore, we illustrate the evolution of the Landau-level spectra as a function of the total layers. The magneto-optical properties and the Coulomb excitations are closely related to the Landau-level spectra and the transition matrix elements that are evaluated from the initial- and final-state wave functions.

\subsection{ Multiple Dirac-cone band structure}
In the absence of external fields, the $N$-layer AA stacked graphene, most symmetric  structure among all the stackings, owns $N$ sets of linear energy subbands intersecting at the K point, each behaving as the monolayer-like Dirac cone\cite{PRB83;165429,JAP110;013725,CPL550;104}. For the odd-layer scenario, one of the N sets is situated close to $E_{F}=0$, mapped to the monolayer, and the other ones are oppositely mirrored to each other with respect to $E_{F}=0$. For the even-layer scenario, half of the Dirac-cone structures are placed above $E_{F}$=0 and the another half below $E_{F}=0$, the mirroring being in accordance with the odd-layer case.
The trilayer case is illustrated to understand the main features of the energy dispersions.
Three sets of subbands are shown in Fig. 10(a) and labeled by indexes $S^{c,v}_{1}$, $S^{c,v}_{2}$ and $S^{c,v}_{3}$.
The Dirac cone structure of $S^{c,v}_{2}$ is nearest to $E_{F}=0$, and those of $S^{c,v}_{1}$  and $S^{c,v}_{3}$ are approximately symmetric to each other.
Similarly, examples of even layers ($N$=2 and $N$=4) are depicted in Fig. 10(b).
It should be noted that the Dirac points at the K point are shifted relatively to
their neighbor by the energy that decreases with the number of layers, as well as
demonstrated in the effective-mass model; the energy difference is given by
$2\cos(j+1\pi/(N+1))$-$2\cos(j\pi/(N+1))$\cite{JAP110;013725}.

\subsection{Landau levels and Landau wave functions}

As it is the general and fewest layer case, we select trilayer graphene to elaborate on the magnetic quantization of the Dirac quasiparticles in the AA stacking configuration.
The LLs at $B_{0}=20$ T can be divided into three groups, as indicated by black, red and blue colors in Fig. 11(a), in which each Landau state at a given ($k_x,k_y$) is four-fold degenerate; the division is based on the energy spectrum and the characteristics of the spatial distribution of the wave function (Fig. 11(b)-11(g)).
In particular, the conduction LLs and valence LLs of each group are approximately symmetric to each other about the zero-mode LL, $n_{1}^{c,v}=0, n_{2}^{c,v}=0$, or $n_{3}^{c,v}=0$, as corresponding to the monolayer-like energy spectrum.

The wave functions, characterized by the subenvelope functions on different sublattices,
are used to precisely define the quantum number of each LL. For the four degenerate Landau states at ($k_{x}=0$, $k_{y}=0$), the wave functions at the 1/6 and 4/6 locations of the enlarged cell have identical amplitudes, as do those at 2/6 and 5/6. Moreover, the $1/6$ localized state has the subenvelope functions $A^{l}_{o}$'s, which are equal to $B^{l}_{o}$'s ($l=1,2$ and 3) of the $2/6$ localized state and vice versa. In Figs. 11(b)-11(g), the subenvelop functions of the $2/6$ localized state are demonstrated, where further discussions are restricted to the odd-indexed components because of the relationship $B^{l}_{o}$=$-B^{l}_{e}$. With the starting energy away from $E_{F}=0$,  the first and third groups of LLs have a monolayer-like relationship between the two subenvelop functions for every layer, while the second group of LLs spreading from $E_{F}=0$ have the same relationship only in the two outmost layers. The subenvelop functions related to the inner layer are empty. This means that the LLs of the second group are entirely regarded as a result of monolayer graphenes, and the Landau quantization of the Dirac cones mapped to the monolayer is consistently presented in the odd-layer cases. On the contrary, all the layers of the even-layer AA-stacked graphenes are equivalent. The quantum numbers of LLs can be identified by the non-empty subenvelop functions of lattice $B$ in any layer.
Using the single-mode wave functions is a visible means to identify the LLs,
since there is an absence of LL couplings under the specific interlayer atomic interactions.
These results indeed reflect the monolayer properties of AA stacking.

The unique spectral features are better understood by investigating the LL evolution
with the field strength $B_{0}$. The three monolayer-like LL spectra of AA-stacked trilayer graphene are shown in Fig. 12.
The energy distribution in each group LL is related to the individual Dirac cones that are primarily separated by the specific interlayer atomic interaction $\alpha_{1}$\cite{PRB83;165429,JAP110;013725,CPL550;104}.
Their energies can be regarded as those of the monolayer LLs, which have a simple square-root dependence on the field strength and the quantum number. Meanwhile, all the LLs directly cross one another, i.e., there is an absence of  LL anticrossings, since each Landau state is characterized by a single-mode harmonic function.

The LL spectra of the AA-stacked bilayer and tetralayer graphenes are shown in Fig. 13 and Fig. 14, which respectively demonstrate the two and four groups of monolayer-like patterns, as well as the oscillating Fermi level.
Each group is roughly symmetric about the zero-mode LL, which is nearly constant as a function of field strength; each level moves towards the individual Dirac point when the magnetic field approaches zero. There are essential differences between the tight-binding model and the effective-mass model. In systems with charge neutrality, the tight-binding model calculates the oscillating Fermi levels based on equivalent numbers of conduction and valence carriers. On the other hand, due to the Dirac cone structures extending to the full k space in the effective-mass model, the isotropic dispersion makes the spectra identical and the Fermi levels consistent with its zero-field value.

\subsection{Magnetic-field-dependent oscillation of the Fermi level}
Around the zero-field value of $E_{F}=0$, the Fermi energy at a constant carrier density oscillates as a function of the field strength, as shown in Fig. 12 by the bold wriggling curves. This leads to an energy gap, the spacing between the highest occupied LL and the lowest unoccupied LL; this gap oscillates with the magnetic field, and results in the metal-semiconductor transition at certain specific fields.
The magneto-resistance measurements to detect the oscillation of the Fermi level were performed in other 2D electron gas systems\cite{PRL62;1173,PRB90;144517,PRB64;201311}.
As for the $B_{0}$-dependent energy gaps in bilayer and trilayer graphenes (black curves in Figs. 15(a) and 15(b)), the oscillation period and amplitude both increase with an increment of the field strength, but decrease in the cases of more layers.
Furthermore, a gate voltage shifts the Dirac cones\cite{CPL550;104}, providing an available means of using external fields to modulate the energy gap.

\subsection{Magneto absorption spectrum}

The main characteristics of the electronic properties is directly reflected in
other physical properties, e.g., optical absorption spectra. In the absence of external fields, previous works show that only intra-subband excitations within the same Dirac-cone structure are permitted, and consequently asymmetric absorption peaks are induced\cite{APL103;041907}. The threshold excitation energies are investigated in the odd-layer and even-layer scenarios. Also shown are their dependence on the stacking layers and, after exceeding certain layer numbers, the cases in both scenarios exhibit critical properties identical to those of 3D graphite for different frequency regions. On the other hand, feature-rich spectra are expected due to the optical transitions between LLs.
There are several kinds of prominent symmetric peaks, resulting from the individual intragroup LL transitions and satisfying the particular selection rule $\Delta n=\pm1$ (same as the monolayer)\cite{APL97;101905}. Transitions between intergroup LLs are forbidden due to the zero electric dipole moments derived from the spatial symmetries of the wave functions in the AA-stacked systems. The effective-mass model demonstrates similar spectra\cite{JAP110;013725}. However, as a result of the identical spectra of groups of LLs and the constant Fermi level, important characteristics were missed in the case of the AA stacking, such as the strength and numbers of peaks and the excitation channels.

\section{Magnetic quantization in AB-stacked graphene}

The AB stacking configuration is stable and commonly observed in nature graphite.
Generally, the AB-stacked graphenes have attracted intense attention due to the fact that the properties of massless Dirac fermions are presented in all cases of an odd number of graphene layers\cite{Rep76;056503,PRB75;193402,PRB78;205425,PRB81;115315,RevModPhys81;109}. Of special interest is the band structure that can be modeled by a hybridization of a monolayer and bilayer graphenes. On the other hand, in all cases of an even number of layers, the energy dispersion is parabolic in the low-energy region.
An evidence of Dirac nature of charge carriers in ultrathin epitaxial graphenes was observed by scanning tunneling spectroscopy\cite{NatMat12;887} and angle-resolved photoemission spectroscopy\cite{PPL98;206802,PRB88;155439,PRL102;056808}.
The Landau quantization gives rise to interesting phenomena of the electronic properties in the even-layer cases of inversion-symmetry graphenes and in the odd-layer cases of symmetry-broken graphenes.

\subsection{Band structure: hybridization with monolayer graphene}
The low-energy electronic properties in AB-stacked bilayer graphene are characterized by four parabolic energy bands, as shown in Fig 16(a): a pair of conduction and valence bands, labeled $S_{1}^{c,v}$ (black curves), and another, labeled $S_{2}^{c,v}$ (red curves).
The first pair of subbands slightly overlaps near $E_{F}=0$ (inset of Fig. 16(a)), and each subband of the second pair is approximately shifted away from $E_{F}=0$ by an energy of $\gamma_{1}$ or $-\gamma_{1}$, where $\gamma_{1}$ is the vertical nearest interlayer atomic interaction. Unlike only parabolic bands in the bilayer graphene, the band structure in a trilayer graphene is hybridized from a monolayer and bilayer band structures, as shown in Fig 16(b). Near the K point, the intersections of linear and parabolic subbands are, respectively, separated, but a slight overlap remains between the valence and conduction bands (inset of Fig. 16(b)). All odd-layer scenarios still exhibit monolayer-like linear dispersions, while all even-layer ones only show parabolic energy dispersions\cite{PRB75;193402,PRB78;205425,Rep76;056503,PRB81;115315,RevModPhys81;109}.
AB-stacked graphenes are gapless 2D semimetals, whereas a tunable energy gap can be induced by applying an electric field in the bilayer case\cite{JPCM18;5849,JPSJ76;024701,
PRB74;161403,PRB75;155115,PRL99;216802}. The theoretical results are experimentally confirmed by infrared spectroscopy \cite{Nature459;820,PRL102;256405}, ARPES\cite{Science313;951} and transport\cite{Nature7;948,NatNanotechnol4;383,
SCIENCE320;206} measurements.

\subsection{Landau levels and Landau wave functions}

The trilayer graphene is used to elaborate on the main characteristics of the magnetic quantization of AB-stacked graphenes. The LLs at $B_{0}=20$ T can be divided into three groups, as indicated by black, red and blue colors in Fig. 17(a).
The first, second and third groups are, respectively, attributed to the quantization of the linear subband $S_{1}^{c,v}$ and  the parabolic subbands $S_{2}^{c,v}$ and $S_{3}^{c,v}$. The LL spacing is smaller in the higher group than in the lower group, a spacing variation determined by the zero-field energy dispersion (DOS). In the first group, the onset energy is close to $E_{F}=0$, and each Landau state has the wave function distributed only on the two outer layers (black curves), which is attributed to the magnetic quantization effect on the linear subbands $S_{1}^{c,v}$. As a result of the degeneracy breaking of the Dirac points (inset of Fig. 16(b)), two double degenerate states are split from the zero mode of the quantized Dirac quasiparticles, with the respective quantum numbers $n_{1}^{c}=0$ (5 meV) and $n_{1}^{c}=0$ ($-$5 meV), being determined by the dominant subenvelope functions of $A_{o}^{1,3}$ (localized at 1/6 and 4/6 unit cell) and $B_{o}^{1,3}$ (localized at 2/6 and 5/6 unit cell), as shown by the black curves in Figs. 17(b)-17(m). The higher conduction LLs are placed with narrower spacings. Their quantum numbers $n_{1}^{c}$ are in an ascending order with an increase of energy, and can be derived from either $A_{o}^{1,3}$ or $B_{o}^{1,3}$ as well.

The second group has two sets of double degenerate LLs, the dominating sublattices of which are $B_{o}^{2}$ for the 1/6 (4/6) localized states and $B_{o}^{1,3}$ for the 2/6 (5/6) localized states, as shown by the red curves in Figs. 17(b)-17(m). The first four double degenerate states, labeled $n_{2}^{v}$=0, 1 and $n_{2}^{c}$=0, 1, are split from the zero and first modes of the $S_2^{c,v}$ subband quantization, with the splitting energies $\sim10-20$ meV. However, the splitting energies for higher LLs are less than 1 meV, which might be difficult to observe in experimental measurements.\cite{PRB88;121302,NatPhys7;621,NatPhys7;953,PRL107;126806} In the same way, the dominating sublattices are $A_{o}^{2}$ for the 1/6 $\&$ 4/6 localized states, and $A_{o}^{1,3}$ for the 2/6 $\&$ 5/6 localized states (blue curves in Figs. 17(b)-17(m)) for the third group. In all, each sublattice equally dominates the magneto-electronic properties.

\subsection{Symmetry-breaking effect on Landau level spectrum}
The LL spectrum of the trilayer graphene demonstrates the monolayer-like LLs in the first group and the bilayer-like LLs in the second and third groups\cite{PRB83;165443,PRB77;155416,PRB84;205448,PRB84;125455}, as shown in Fig. 18. The onset energies of the conduction LLs in the first, second and third groups are identified as the band-edge state energies of subbands $S_{1}^{c}$, $S_{2}^{c}$ and $S_{3}^{c}$, respectively, and the same identification is applied to the valence LLs.
For the first group, the energy difference between the onset valence and conduction LLs is equal to the energy splitting of the Dirac points at the K point, and for the second group it is determined by the gap between the $S_{2}^{v}$ and $S_{2}^{c}$ parabolic subbands. %
Being split from $E_{F}=0$, the 12 ($4N$) Dirac-point LLs, including the two double degenerate ones of $n_{1}^{c,v}=0$ and the four double degenerate ones of $n_{2}^{c,v}=0$ and $n_{2}^{c,v}=1$, are confined in a narrow energy range ($\sim$ 10 meV) and insensitive to the field strength. Besides, the splittings between the 1/6 (4/6) and 2/6 (5/6) localized states are revealed for the low-lying LLs. The intergroup LLs frequently cross one another, but the anticrossings between them only occasionally appear in strong fields. The quantum transport experiments have verified the degeneracy change of the low-lying LLs due to the lack of inversion symmetry and the LL crossings in AB-stacked trilayer graphenes\cite{NatPhys7;621}.

Essentially, the LL spectra of AB-stacked graphenes can be classified into two kinds based on whether the numbers of their stacking layers are even or odd \cite{PRL96;086805,PRB77;085426,PRB83;165443,PRB77;155416,PRB84;205448,PRB84;125455}. To comprehend the general results, we illustrate the spectra of even-layer AB-stacked graphenes. Unlike the degeneracy breaking in the trilayer graphene, the LL spectra of even-layer AB-stacked graphenes all consist of four-fold degenerate LLs due to the inversion symmetry of the geometric structures, such as the spectra of bilayer and tetralayer graphenes shown in Fig. 19 and Fig. 20, respectively.
The field-dependent energies depend on the subband dispersions in a zero field.
One can find the dominant sublattice for each group, i.e. in the ascending sequence of groups followed by $B_{o}^{1}$ ($B_{o}^{2}$) and $A_{o}^{1}$ ($A_{o}^{2}$) for bilayer graphene\cite{PRB77;085426}, and $B_{o}^{2}$ ($B_{o}^{3}$), $B_{o}^{1}$ ($B_{o}^{4}$), $A_{o}^{1}$ ($A_{o}^{4}$), and $A_{o}^{2}$ ($A_{o}^{3}$) for tetralayer graphene\cite{PRB90;205434}. The $4N$ Dirac-point related LLs, insensitive to magnetic fields, are confined in the vicinity of $E_{F}$=0, and classified as $n_{1}^{v}$=0 and $n_{1}^{c}$= 0 for bilayer graphene, or $n_{1}^{v}$=0, $n_{1}^{v}$=1, $n_{2}^{c}$=0, and $n_{2}^{c}$=1 for tetralayer graphene.

As a result, it is deduced that for the LLs in $N$-layer graphene, there are $[(N+1)/2]$ groups starting to appear near $E_{F}=0$, of which one especially exhibits a monolayer-like energy spectrum once $N$ is odd, while the other $N-[(N+1)/2]$ groups start to appear away from $E_{F}=0$. The degeneracy of the LLs is broken in the odd-layer cases, while it is preserved in the even-layer cases due to the inversion symmetry. Furthermore, the anticrossing of LLs between two neighboring groups occurs in the tetralayer graphene at certain regions of magnetic field strengths and energies, as the ellipses depict in Fig. 20.
It is revealed that the interlayer atomic interaction $\gamma_{3}$ couples the LLs in the 3 modulo\cite{PRB83;165443,PRB87;075417,JPSJ17;808}, so that according to the Wigner-von Neumann non-crossing rule, two multi-mode LLs containing identical modes do not cross each other as a function of the magnetic field strength. An explanation is made in detail in the following chapter 6.3. With an increasing number of layers, the LL anticrossings are easily observed. While experimental observations on QHE confirm the low-energy spectra through the specific sequence of Hall conductivity plateaus in the AB-stacked bilayer\cite{NatPhys2;177,NatMater7;151}
and trilayer graphenes\cite{NatPhys7;621,PRX2;011004,NanoLett13;1627}, their dependences in multilayer systems need to be systematically verified.

The dominating sublattices for each subband under magnetic quantization obey a regular rule. For odd-layer graphenes, the dominating sublattices for the first group of LLs quantized from the linear subbands are the two outmost layer sublattices $A_{o}^{1}$, $A_{o}^{N}$ or $B_{o}^{1}$, $B_{o}^{N}$. The other $[\frac{N+1}{2}]$-1 ($N-[\frac{N+1}{2}]$) groups starting from (away from) $E_{F}=0$ are followed by sublattices $B$ ($A$) being counted from the inner (outer) to the outer (inner) layers in the ascending sequence of  groups. In the same way, even-layer graphenes own a rule similar to the former, but excluding the linear-band part. This generalization for the magneto-electronic properties is valuable in understanding other physical properties, such as the mechanisms of magneto-optical properties and Coulomb excitations, etc.

\subsection{Magneto absorption spectrum}
The optical absorption spectra reflect the main characteristics of the electronic properties.
In the absence of fields, only inter-subband excitations between the same pairs of parabolic subbands induce asymmetric absorption peaks\cite{JAP108;043509}; those excitations can be enhanced by applying an electric field\cite{PRB73;144427}. A magnetic field, on the other hand, induces symmetric prominent delta-function-like peaks as a result of the intra- and inter-group LL transitions \cite{PRB77;115313}. According to the Fermi's golden rule, the former transitions satisfies the monolayer selection rule $\Delta n=\pm1$, while the latter follows more complicated selection rules, which are not unified for all inter-group transitions and should be identified by checking the probability of the dipole transition. In a previous work\cite{PRB77;115313}, an AB-stacked bilayer system is shown to exhibit an absorption spectrum consisting of four types of absorption peaks, with two being the result of intra-group transitions and the other two being the result from inter-group transitions. Each type of optical excitations lead to twin-peak structure owing to the asymmetry of valence and conduction LLs about the Fermi level (electron-hole asymmetry). The inter-group selection rules are $\Delta n_{12}=0$, $\Delta n_{12}=-2$, $\Delta n_{21}=0$, and $\Delta n_{21}=2$ where the subscript index represents the group number. It is evident that the two different rules are both based on the underlying concept that the initial and final states in the dipole transition have the same zero points in subenvelop functions. This concept provides an insight into multilayer systems. The specific selection rules for any inter-group transitions can be straightforwardly derived from the wave functions. It should be noted that for the odd-layer cases, the excitations from the quantized LLs of the linear bands to those of the parabolic bands are forbidden due to the zero dipole transition moment. Therefore, $N$-layer AB-stacked graphene is deduced to have a total of $N^{2}$ types of absorption peaks, resulting from $N$ intra- and ($N^{2}-N$) inter-group transitions for even-layer cases; a total of ($(N-1)^{2}+1$) types including $N$ intra- and ($(N-1)^{2}-(N-1)$) inter-group transitions for odd-layer cases. The twin-peak structures are presented in recent magneto-absorption experiments\cite{PRB83;125302,PRB85;245410}, where the measured transition energies, however, are limited to few hundreds of meV. The magneto-optical properties need to be verified in detail over the entire spectrum range within the consideration of intra- and inter-group LL transitions. 

\subsection{Magneto plasmons}

In addition to optical properties, electronic excitations are also enriched by the magnetic quantization. The complicated Coulomb screening in multilayer graphene can be characterized by the dielectric-function matrix\cite{PRB74;085406,PLA352;446}. The well-developed layer-dependent matrix is based on the tight-binding functions of distinct sublattices, as done for the magnetic Hamiltonian matrix in the generalized tight-binding model. As a result, the interlayer e-e interactions, the interlayer atomic interactions and the magnetic field are simultaneously included in the calculations of Coulomb excitation spectra. By utilizing this matrix, the screened energy-loss function is obtained, which is then used to determine the main features of collective Coulomb excitations.  The previous study shows that monolayer graphene exhibits a lot of low-frequency  magneto-plasmons at zero temperature even if it is a zero-gap semiconductor\cite{ACSnano5;1026}. Each magneto-plasmon mode is associated with a specific inter-LL  excitation channel from an occupied LL to another unoccupied one. Magneto-plasmons, quanta of the electron density oscillations in a perpendicular magnetic field, present the novel dependence in the relationship between their frequencies and the transferred momentum, since the longitudinal Coulomb force strongly competes with the transverse Lorentz force.
The intensity, frequency, and number of magneto-plasmons are significantly affected by the transferred momentum, the magnetic-field strength, the temperature and the doping carrier density\cite{PRB85;235444,PRB80;085408,PRB84;035439,PRB84;115420,PRB75;115314,
JAP109;113721}.

Such characteristics are further diversified by the distinct stacking configurations. For AB-stacked and AA-stacked bilayer graphenes, there exist complicated energy-loss excitation spectra with more inter-LL plasmon resonance peaks\cite{PRB89;165407}. In the former the energy-loss spectrum is mainly dominated by discrete inter-LL excitations, while in the latter a 2D-like plasmon involving the entire low-frequency Landau states is found. Also noticed that a 2D-like plasmon, being similar to that of a 2D electron gas in the absence of magnetic field, is a result of the highly symmetric AA stacking and the dense LL distribution around the Fermi level. The inter-LL plasmon and the 2D-like plasmon have a very different dependence on the transferred momentum and the magnetic field strength. These two kinds of magneto-plasmons are expected to be drastically changed by the increasing layer number. Whether there are new collective excitation modes in ABC-stacked graphenes deserves a further detailed study.

\section{ Magnetic quantization in ABC-stacked graphene}

The ABC stacking configuration stably exists in nature graphite\cite{PRSLSA106;749,PRSLSA181;101}. The surface-localized states of ABC-stacked graphenes form partially flat bands near the Fermi level\cite{PRL73;245426,PRB83;220503,PRB87;155116}, giving rise to an enhancement of the density of states of LLs. A prominent peak appears at the Fermi level with the intensity proportional to the layer numbers as a result of the increasing LLs that originate from the Dirac points. In particular, the band structure exhibits sombrero-shaped energy dispersion around the energies of the nearest vertical intralayer hopping integral\cite{RevModPhys81;109,PRL97;036803,PRB84;165404,PRB82;035409}. Consequently, the Landau-level spectrum reveals intriguing features within such a band width\cite{PRB90;205434}: LLs with energies inversely proportional to the magnetic field strength and intra-group Landau-level anticrossings in the region of weak magnetic fields.

\subsection{Band structure: surface-localized states and sombrero-shaped energy bands}

The low-energy electronic properties of ABC-stacked trilayer graphene are characterized by three pairs of conduction and valence subbands, $S_{1}^{c,v}$, $S_{2}^{c,v}$ and $S_{3}^{c,v}$, displaying partially flat, sombrero-shaped and linear shaped dispersions, respectively, as shown in Fig. 21(a).
In the vicinity of $E_{F}=0$, the subbands of the first pair $S_{1}^{c,v}$, are weakly dispersive near the $K$ point. As a result of the trigonal warping effect \cite{Ann326;721,PRB80;165409}, the low-energy dispersions are anisotropic with three crossing points at $E_{F}=0$, which are identified as Dirac points\cite{PRB80;165409}, i.e., one in the $K\Gamma$ direction and another two rotated with respect to the former by $120^{o}$ and $-120^{o}$ (inset of Fig. 21(a)). The states in this region are formed predominantly by atomic orbitals on the two outermost layers, known as surface localized states \cite{PRL73;245426,PRB83;220503,PRB87;155116}.
Recent experiments confirm that a tunable energy gap can be induced by applying an electric field in the trilayer cases\cite{PRB87;165102,Nature7;948,NatNanotechnol4;383}. Away from $E_{F}=0$, the conduction subbands and the valence subbands, respectively, cross near the energies of  $\beta_{1}$ (=0.29 eV) and $-\beta_{1}$ (=-0.29 eV), corresponding to the band edges of the subbands $S_{2}^{c,v}$ and $S_{3}^{c,v}$. In particular, the sombrero-shaped conduction (valence) subband, $S_{2}^{c}$ $(S_{2}^{v})$, has a local energy minimum (maximum) and maximum (minimum), measured respectively, as 0.29 (-0.29) eV and 0.36 (-0.36) eV, within which two closed circular loops at a given energy lead to different quantized Landau orbitals.

Tetralayer stacked graphene has a similar band structure, except for one more pair of parabolic bands located near $\pm\beta_{1}$, as shown in Fig. 21(b). Surface localized states are distributed around $E_{F}=0$ while four Dirac points come into existence, as shown in the inset of Fig. 21(b). The theoretical band structure is experimentally confirmed by infrared absorption spectroscopy\cite{PRL104;176404}. For all ABC-stacked graphenes, the energy dispersion of the surface localized states is simply described by a relationship $E \propto k^{N}$ in the minimal model\cite{RevModPhys81;109,PRB82;035409,PRL73;245426,PRB80;165409}. Furthermore, it is shown that the number and characteristics of the Dirac points are related to the number of the stacking layers\cite{RevModPhys81;109,PRB80;165409,PRB90;205434}. In addition, more of the subbands intersecting near $\pm\gamma_{1}$ show a sombrero-shaped energy dispersion with the increasing number of layers. The feature-rich band structure leads to the intricate and versatile magnetic quantization in ABC-stacked graphenes.

\subsection{Landau levels and Landau wave functions}

The magnetically quantized LLs of ABC-stacked trilayer graphene subjected to $B_{0}=25$ T are shown in Fig. 22(a).
The LLs can be divided into three groups according to the spatial characteristics of the wave functions, as indicated by black, red, and blue colors in Figs. 22(b)-22(g). The first, second and third groups, respectively, result from the quantization of the weakly dispersive $S_{1}^{c,v}$, the sombrero-shaped $S_{2}^{c,v}$, and the linear $S_{3}^{c,v}$ subbands. In particular, the onset energies and level spacings for all groups are relevant to the zero-field energy dispersions. Each level is four-fold degenerate due to the inversion symmetry, leading to the $1/6$ ($4/6$) localized state having the subenvelop functions $A^{l}_{o}$'s ($l=1,2$ and 3 for layer numbers) equal to $-B^{l}_{o}$'s ($l=3,2$ and 1) of the $2/6$ ($5/6$) localized state and vice versa.

The first group starts to form near $E_{F}=0$, in the vicinity of which the three degenerate LLs in the minimal model are slightly split by remote interlayer hoppings\cite{Ann326;721,PRB80;165409,PRB90;205434}. Away from $E_{F}=0$ the spacing becomes more obvious and the LLs are clustered together with increasing energy due to the higher DOS of the $S_{1}^{c}$ subband. As evidenced by the dominating subenvelop function $B^{1}_{o}$ in Fig. 22 (c), the LLs are counted by increasing quantum numbers with increasing energies. On the contrary, the quantization related to the opposite outmost sites of $A^{3}$ is introduced to the $1/6$ and $4/6$ localized states. Near $\beta_{1}$, the LLs are densely packed as a consequence of the crossover of the $S_{2}^{c}$ and $S_{3}^{c}$ subbands, and the second and third group LLs appear in the sequence of the quantum numbers determined by the subenvelop functions $B^{3}_{o}$ and  $B^{2}_{o}$ in Figs. 22 (g) and 22(e).

\subsection{Intra- and inter-group Landau-level anticrossings in the field-dependent energy spectrum}

The three groups of LLs display feature-rich  energy spectra during a variation of field strength, as shown in Fig. 23.
Each LL moves towards the respective onset energy for every group, i.e., the local extreme value near the K point (Fig. 21(a)) with respect to each subband in the absence of an external field. The three Dirac-point related LLs are insensitive to the variation of field strength and confined in a small energy range near the Fermi level, with their separated energies mainly determined by the specific interlayer atomic interaction $\gamma_{4}$. In contrast to what the monotonically B-dependent energies present in the first and third groups, a complicated pattern of non-monotonic LL dispersions is observed in the second group under certain critical fields. This is responsible for the particular magnetic quantization rule relating to the sombrero-shaped energy dispersion.
There are several inverted LLs located between energies of the local maximum and minimum of this sombrero-shaped subband. The relationship between each LL and the critical magnetic field is determined by the interplay between the magnetic degeneracy and the total number of states enclosed in this area. A more detailed discussion is made as follows.

The LL spectrum exhibits special phenomena, the intragroup and intergroup LL anticrossings, the exploration of which is performed by following the evolution of the wave functions. It is seen that two LLs repel each other, as in the region of the dashed ellipses in Fig. 24, when these two levels are in modulo 3\cite{PRB83;165443,RSCAdv4;56552,PRB90;205434,JPSJ17;808}.
This event can even be triggered among all different groups in a sufficiently strong magnetic field. This is a result of the Wigner-von Neumann non-crossing rule, which states that two multi-mode LLs do not cross as a function of the magnetic field strength if they simultaneously have identical modes. From a qualitative perturbation analysis, it is deduced that there are three factors, $\beta_{2}$, $\beta_{3}$ and $\beta_{5}$, that induce the LL anticrossings in ABC-stacked graphenes. The energy perturbation caused by $\beta_{2}$ has its expectation value between the two same mode subenvelop functions $B^{1}$ and $A^{3}$ of the two hybridized states. A likewise explanation of the $\beta_{3}$-induced correction is applicable to $\beta_{5}$, which is also responsible for the specific relationship between $B^{1}$ and $B^{3}$, and between $A^{1}$ and $A^{3}$. The drastic changes of the wave functions during the anticrossing processes, e.g., for $n_{2}^{c}=2$ and $n_{3}^{c}=0$ LLs, are illustrated in Figs. 24(b)-24(g). In principle, for Landau states away from the anticrossing regions (the first five circles Fig. 24(a)), the two subenvelop functions established on the perpendicular projected sublattices in adjacent layers have the same number of the zero points; two sublattices in the same layer have zero-point numbers differed by one. As shown in the lower part of Fig. 24(b)-24(g), the non-perturbed $n_{2}^{c}=2$ LL has well-defined subenvelop functions that have zero points of 3, 4, 2, 3, 1 and 2 in the order of $\{A_{o}^{1}, B_{o}^{1}, A_{o}^{2}, B_{o}^{2}, A_{o}^{3}, B_{o}^{3}\}$. However, at the center of the anticrossing region, depicted in the middle part of Figs. 24(b)-24(g) ($\sim$0.37-0.41 eV), the perturbed subenvelop functions possess multi modes composed of the main mode $n_{2}^{c}=2$ and the side mode $n_{3}^{c}=0$.
This significant hybridization results in the two components being comparable, while away from this region, the wave function converts to the characteristics of the well-defined $n_{3}^{c}=0$ Landau state (upper part of Figs. 24(b)-24(g)) as a result of the decreasing perturbation.

A further exploration of the different layer cases is worthwhile in order to comprehend the unusually sequenced LLs and the LL anticrossings. Except for one pair of bands contributed by the surface localized states that is always weakly dispersive near the Fermi level, there are $N-1$ pairs of conduction and valence subbands intersecting in the vicinity of the energies $\pm\beta_{1}$ \cite{PRB84;165404}. Some of them display the sombrero-shaped energy dispersions which determine the unique pattern in the LL spectrum.
Tetralayer graphene shows only one group of LLs inversely dependent on the field strength in Fig. 25, and pentalayer graphene shows two such groups in Fig. 26.

However, a relatively narrow region of the inverted LLs is observed for the graphene with fewer layers, which means that the enclosed states in the sombrero-shaped band structure are relatively small, as are the corresponding critical magnetic fields $B_{s}$. It is deduced that with an increase of stacking layers, $B_{s}$ becomes stronger and the intragroup LL anticrossings show up in a relatively large energy range as a result of more sombrero-shaped subbands\cite{PRB90;205434}. The derived rules are that the number of groups showing self anticrossing is, respectively, $(N-1)/2$ and $(N-2)/2$ for odd and even $N$. Moreover, the generalization of the Dirac-point related LLs separated near $E_{F}=0$ is also extended to the $N$-layer systems, as shown in Fig. 27.
There exist $N$ four-fold degenerate LLs with the energy distribution insensitive to the magnetic field and limited in a very narrow region evaluated to be $\sim10$ meV for $N\gtrsim10$. The aforementioned results reveal feature-rich spectra in the cases of more layers.

\subsection{Peculiar quantization rule for sombrero-shaped energy bands}

\subsubsection{Inverse dependence of Landau-level energies on the magnetic field}

As a signature of the ABC-stacked graphenes, the intragroup LL anticrossings frequently appear in an energy region within the local maximum and the local minimum of the sombrero-shaped subband, as shown in Fig. 28(a), especially for weak fields where the inverted LLs is densely sequenced.
We consider the Hamiltonian without the anticrossing interaction in order to clearly realize the particular quantization rule for the sombrero-shaped subband. The inverted curvature of such a subband structure gives rise to two available Landau orbitals at a given energy, one located at the inner circular loop and the other at the outer circular loop of the sombrero-shaped subband surface. A weak magnetic field aggregates only the states near the vertex of the inner surface, so that the LL energies at first decrease with increasing field strength as more inner-part states are quantized into LLs. At a critical magnetic field, labeled $B_{s}(n_{s})$,
the LL with the quantum number $n_{s}$ reaches the minimum of its energy distributions, which is related to the energy minimum of the sombrero-shaped subband, and then converts back to the typical $B_{0}$-dependence with further increasing field strength. This means that under the critical field $B_{s}(n_{s})$, all the electronic states covered by the inner sombrero-shaped surface are magnetically quantized to the inverted LLs, in which the LLs  are entirely filled up to the quantum number $n_{s}$. In contrast, the quantization of the states on the outer surface is introduced to the $B_{0}$-proportional LLs with quantum numbers larger than $n_{s}$.

\subsubsection{Periodical quantization effect related to the inverse of the critical magnetic field}

Given that the magnetic degeneracy of each LL is $D=(eL^{2}/2\pi\hbar c)=\rho B$, one can derive the relation $N_{s}=n_{s}\times D$ at the critical point, where $N_{s}$ is the number of total states on the inner surface.
At a stronger magnetic field, fewer LLs can occupy the totally enclosed states due to the higher degeneracy $D$, and therefore the number of the unusual LLs decreases.
Once the magnetic field is stronger than $B_{s}(n_{s}=1)$, a series of all the LLs evolves into an ascending sequence with increasing energy. The relationship between $B_{s}$ and $n_{s}$ is simply characterized by $B_{s}\times n_{s}=$ const, as indicated by the constant $\triangle\,(1/B_0)$ period in Fig. 28(b). Therefore the total degeneracy $D\times n_{s}$ equals the states enclosed in this area, i.e., $N_{s}=D\times n_{s}\propto B_{s}\times n_{s}$, that is to say, the inversely proportional relationship between $B_{s}$ and $n_{s}$ characterizes the magneto-electronic properties resulting the sombrero-shaped band structure.

\subsection{Magneto absorption spectrum}
In the absence of external fields, optical absorption spectra exhibit shoulder and peak structures in the low-frequency spectrum, dominated by the excitations from the surface localized states\cite{SynMet162;800,PRB83;245418}. The absorption frequencies and intensities can be modulated by an applied electric field\cite{JPSJ76;024701}. On the other hand, symmetric delta-function-like peaks are expected to be presented in the magneto absorption spectra as a result of the direct intra- and inter-group LL transitions. The particular selection rule $\Delta n=\pm1$ is applied to the individual intragroup LL transitions. However, alike to the manner of the AB stacking, the allowed inter-group LL transitions are based on the underlying concept that in the dipole transition, the initial and final states have the same zero points in the subenvelop functions.
For an $N$-layer system, there are $N^{2}$ types of absorption peaks deduced from the transitions of which a total of $N$ are from intra groups and a total of ($N^{2}-N$) are from inter groups. Another interesting aspect of the absorption spectrum is that the LL anticrossing effect induces extra selection rules of modulo 3 for the intra-group and inter-group transitions. This is especially important for the intra-second group transitions under magnetic fields smaller than the critical strengths\cite{PRB90;205434}, where the frequently appearing anticrossings lead to a considerable number of peaks with intensities proportional to the LL hybridization.

\section{ Differences among AA-, AB- and ABC-stacked graphenes }

\subsection{Landau-level spectrum}

The stacking configuration plays an important role in the electronic properties of multilayer graphenes. Of interest is the respective energy band structures and the corresponding LL spectra of different stacking configurations, i.e., AA, AB and ABC stackings, as reviewed in the Chapters. 3-6. In the absence of external fields, the AA-stacked graphenes preserve the conical Dirac cone structures due to the vertically projected geometric structure. On the contrary, the AB- and ABC-stacked graphenes display rather distinct band structures, with significant hybridization caused by the interlayer atomic interactions. The band structure of AB-stacked graphene consists of several bilayer-like parabolic bands, while particularly in odd-layer cases, it owns additional monolayer-like linear bands. In ABC systems, there is always one pair of weakly dispersive energy bands near the Fermi level; those bands are formed by surface localized states.
In addition, some bands peculiarly display sombrero-shaped dispersions near the energy of the vertical nearest-neighbor interlayer interaction. The number of the sombrero-shaped subbands is increased when the number of layers is increased. For both AB- and ABC-stacked graphenes, several Dirac points are located at low energies with their characteristics related to the stacking configuration and the layer number.

The zero-field band structures determine the main characteristics of the quantized LL states, such as the sublattice dominance in Landau states, the field-dependent LL energies and the inter- and intra-group LL anticrossings. Each group of LLs is responsible for the magnetic quantization of the respective subband. The $B_{0}$-dependent energies and the onset energies of the LLs are determined by the energy dispersions and the local extreme values of the subband. According to the sublattice dominance for each group, the quantum numbers are well defined based on the traits of the Landau wave functions. The LLs are four-fold degenerate in both AA and ABC systems, while the degeneracy is broken in odd-layer AB systems due to the lack of inversion symmetry. The broken degeneracy is demonstrated by the Landau wave functions with different localization centers.
The $B_{0}$-insensitive LLs are presented for all the three stacking configurations, reflecting the properties of the Dirac points.
However, unlike the large energy separations among them in AA systems, the distribution of such LLs is smaller than 10 meV for AB and ABC systems, with the slight LL splittings mainly caused by the interlayer atomic interaction $\beta_{4}$ ($\gamma_{4}$).

The intra- and inter-group LL anticrossings are important in detailing the magneto-electronic properties of multilayer graphenes. This can be attributed to the Wigner-von Neumann non-crossing rule which prevents the coupled LLs in the modulo 3 from crossing one another during a variation of the field strength. The AA-stacked graphenes present direct crossings by either the same or different groups since there exist no couplings among LLs so that each LL is described by a single mode. While the inter-group anticrossings occasionally occur only between two neighboring groups for AB-stacked graphenes, they can be triggered among all groups for ABC-stacked graphenes. On the other hand, a signature characteristic of the ABC system is the appearance of the intra-group LL anticrossings in the energy region in which the LLs constituted by the quantized states on the sombrero-shaped subband surface densely appear. The low-lying magneto-electronic spectrum can be verified by STS and QHE experiments, while the confirmation of the entire spectrum can be achieved through optical spectroscopy measurements over a wide energy range, including absorption peaks, spectral widths and intensities of different optical transitions, as in Chapters 4.4, 5.5, and 6.5.

\subsection{Wave functions and density of states of Landau levels}
The density of states (DOS), reflecting the main characteristics of of the LL spectrum, is defined as
$D(\omega )=\underset{n^{c},n^{v}}{\sum}{ \int
_{1stBZ}\delta (\omega-E^{c,v}(n,k))dk}.$
The $B_{0}$-dependent DOS exhibits many symmetric delta-function-like peaks attributed to the discrete LLs. The LL energies can be verified by the energies of the sharp peaks in the differential conductance map of STS ($dI/dV$ versus $V$), where the tunneling conductance ($dI/dV$) is proportional to the DOS at the tunneling electron energy\cite{SCIENCE324;924,NATURE467;185,NATURE7;245,PRL109;116802,PRL109;176804,
NATURE7;649}.
For the DOS of the AA-stacked trilayer graphene, the energies of the zeroth LLs of the first, second and third groups are, respectively, evaluated as $-$0.47, 0.01 and 0.46 eV (black, red and blue curves), as shown in Fig. 29(a).
The peaks of every groups are located at energies followed by the relationship of monolayer graphene, i.e., $E^{c,v}\propto\sqrt{n^{c,v}B}$.
A crossover of LLs in different groups leads to higher DOS and stronger tunneling currents in experimental measurements.

In AB-stacked trilayer graphene, the first group is followed by the square-root dependent sequence of monolayer graphene, and the second and third groups correspond to AB-stacked bilayer graphene, as shown in the Fig. 29(b). With half intensity, the peaks near the Fermi level are attributed to the symmetry-broken structure of AB-stacked trilayer graphene that leads to a considerable splitting ($\sim$10 meV) of the zeroth LLs in the first and second groups, while the spillting is hardly observable for the higher LLs under this field strength.
On the contrary, the peak sequence of ABC-stacked trilayer graphene is neither like that of monolayer or bilayer graphenes, as shown in the Fig. 29(c). The first peak at the Fermi level, with intensity estimated to be three times than others, is a composition of three peaks originating from the Dirac points. Its intensity is proportionally increased with an increase of the number of graphene layers. On the other hand, the onset energies for the second and third groups are, respectively, evaluated as 0.34 eV and 0.40 eV, which approach to the crossover of the $S_{2}^{c}$ and $S_{3}^{c}$ subbands under a weak magnetic field. Essential differences of DOS are the spacings and heights of peaks, which can be verified by STS.

In addition, the real-space wave functions of LLs are observed by spectroscopic-imaging STM in 2D electron gas and topological insulator. From the measurements on the variations of the local DOS in graphene planes, spectroscopic-imaging STM reveals the nodal structures of the wave functions attributed to the well-behaved LLs, where the zero-point numbers of the wave functions are identified by fitting the Fourier transformation (FT) of the LDOS data to the FT of the Landau wave functions. For the hybridized wave functions in the LL anticrossing regions, e.g., in AB- and ABC-stacked graphenes \cite{RSCAdv4;56552,PRB90;205434,PRB83;165443}, spectroscopic-imaging STM provides a way to distinguish the main mode and the side modes of the LLs. Also, it performs the verification on the LL splitting resulting from the symmetry broken down in AB-stacked graphene \cite{PRB84;161406,RSCAdv4;56552,PRB84;075451,
PRB87;085424,PRB87;115422,PRB87;075417}. STS and STM directly measure the energy spectrum and the internal structure of wave functions, giving a useful identifiable picture of the Landau levels. The experimental results can be used as a tool to experimentally identify the stacking configuration of a multilayer graphene.

\section{Concluding remarks }

The generalized tight-binding method is widely used to study the physical properties of massless and massive Dirac fermions in graphene-related systems. We review the magneto-electronic properties of multilayer graphenes, focusing on the energy spectra and the related subenvelope functions in the AA, AB and ABC stacking configurations.
Depending on the layer number, stacking configuration and geometric symmetry, the LL spectra show feature-rich characteristics, including novel quantum numbers of the initial LLs, versatile field-dependent energies, high state degeneracy, semiconductor-metal transitions and inter- and intra-group anticrossings. Based on the sublattice dominance, the LLs of an $N$-layer graphene can be clearly classified into $N$ groups, with the quantum numbers identifiable by the derived dominant subenvelop functions.
The degeneracy of LLs is broken by applying an external gate voltage to the graphene sheet. The onset energy and the field-dependent LL energies of each group are relevant to the zero-field band structure. The relationship among subenvelop functions of distinct sublattices is specific for a LL under a given stacking configuration. Via a perturbation theory, we can realize evolution of the subenvelop functions during the inter- and intra-group anticrossings. Few-layer systems have a rich variety of spectra with respect to different number of layers and stacking configurations. Optical spectroscopy, STM and quantum Hall transport provide accuracy in measuring the low-lying magneto-electronic properties. The coupling of the inter- and intra-group anticrossing LLs leads to unexpected physical properties, such as the extra absorption peaks and optical selection rules, and the variation of QHE conductivity plateaus and VHS in the DOS. In addition, spectroscopic-imaging STM can be used to distinguish the main mode and side modes of the hybridized LLs by measuring the variations of the LDOS.

Magnetic quantization in AA-stacked multilayer graphenes results in the square-root LL energy dependence on the field strength and the quantum number, the monolayer characteristic, since the Dirac cones derived from the theoretical calculations and experiments, such as STS and ARPES, are preserved at the hexagonal corners, i.e., the $K$ and $K^{'}$ points. For monolayer, the square-root dependence is limited to energies smaller than 0.4 $\gamma_{0}$, while with respect to each group in a multilayer case, the energy distribution is determined by whether the number of layers is odd or even. If the total number of layers $N$ is odd, there is only one group situated close to the Fermi level, and other $N-1$ groups are oppositely mirrored to each other with respect to the Fermi level. The Landau quantization mapped to the monolayer graphene is consistently presented once $N$ is odd. In particular, the subenvelop functions with non-empty distribution have an identical relationship only for every other adjacent layer. In the even-$N$ case, half the groups are above $E_{F}$=0 and the other half lies below $E_{F}=0$ with the mirror distribution described in accordance with the odd-$N$ case;
all layers are equivalent here. The quantum numbers for either case are identifiable by the non-empty subenvelop functions of lattice $A$ ($B$) of any layer. As a result, the zero-mode LLs correspond to the Dirac points and their energies are insensitive to the field strength. The absence of LL anticrossings in the energy spectra is consistent with the fact that we can use a single mode to identify each LL due to the lack of coupling among LLs in the AA stacked configuration.

AB- and ABC-stacked graphenes both show a non-monotonic LL energy dependence and reveal a complex pattern of LL anticrossings due to the geometric symmetry.
There is a total number of 4$N$ Dirac-point LLs distributed over a narrow energy width ($\sim$10 meV) near $E_{F}=0$. The LL spectrum of an $N$-layer AB-stacked graphene consists of $[(N + 1)/2]$ groups starting to form near $E_{F}=0$ and another $N-[(N + 1)/2]$ groups away from $E_{F}=0$. Once $N$ is odd, one group of LLs especially exhibits a monolayer-like energy spectrum, in which the Landau states are contributed by electronic states localized on every other adjacent layer. The survival of monolayer properties is attributed to the Landau quantization of the massless Dirac quasiparticles.
The degeneracy of LLs is broken in the odd-layer case, while it is protected in the even-layer case due to the inversion symmetry. Quantum transport experiments have verified the variations in state degeneracies caused by the LL crossing and asymmetry-induced breaking through measurements of the Hall conductance plateaus.
The dominating sublattices for energy subbands under magnetic quantization obey a regular rule. The two outermost layer sublattices are for the monolayer-like LLs, and the other different inner layer sublattices $B (A)$ are for the other groups starting near $E_{F}=0$ (away from $E_{F}=0$). Furthermore, the anticrossing of LLs is attributed to the fact that the interlayer atomic interaction $\gamma_{3}$ couples the LLs in the 3 modulo, which prevents multi-mode LLs with certain identical modes from crossing each other in the $B_{0}$-dependent energy spectrum.

For ABC-stacked graphenes, a versatility of magneto-electronic properties is observed when the number of layers is increased. With an onset energy near $E_{F}=0$, there is always one group of LLs responsible for the magnetic quantization of the surface localized states. The phenomenon of LL anticrossings is strikingly pronounced in ABC-stacked graphenes, in contrast to the few anticrossings observed in AB-stacked graphenes and the absence of anticrossings in AA-stacked graphenes. In the energy spectrum, the LL evolution under a magnetic field reveals a complex pattern of two kinds of anticrossings by intergroup and intragroup LLs, and some peculiar LLs attributed to the Landau quantization of the sombrero-shaped energy subbands. The intergroup anticrossings appear for any two groups under a sufficiently strong magnetic field. On the other hand, intragroup LLs frequently anti-cross each other in a small energy region where the peculiar LLs, with energies inversely dependent on $B_{0}$-field strength, are arranged in the order of decreasing quantum numbers when the energy is increased. However, the unusually sequenced LLs and the intragroup anticrossings start to disappear at a critical field strength, in which $B_{C}$ is higher for a system with more layers.
In addition to the nonvertical hopping integral $\beta_{3}$ between two adjacent layers, the two hopping integrals $\beta_{2}$ and $\beta_{5}$ between two next-neighboring layers also take the same role and combine the LLs in the 3 modulo in ABC-stacked graphenes. This is attributed to the specific relationship between the subenvelop functions and the interlayer atomic interactions.

A generalization of magneto-electronic properties is valuable in understanding other physical properties, such as the mechanisms behind magneto-optical excitations.
The intragroup transitions between well-defined LLs satisfy the particular selection rule $\Delta n=\pm1$ in the AA, AB and ABC stacking configurations, whereas the rule for intergroup transitions depends on the configurations.
The former is also applicable for the LLs of Dirac quasi-particles in the Graphene-like 2D materials, e.g., MoS$_{2}$\cite{PRB89;155316,APLHo;acc} and Silicene \cite{PRB89;155316,PRL110;197402} and the topological insulator \cite{PRB85;195440,PRB81;125120,PRB82;045122,APL100;161602}.
The intergroup transitions are forbidden due to the derived zero electric dipole moment in the AA-stacked system. Also, for the AB-stacked graphenes with odd number of stacked layers, excitations from the quantized LLs of the linear bands to those of the parabolic bands are forbidden.
However, considering intergroup transitions for the other cases, the optical selection rules are defined by the concept of dipole transition. Furthermore, the hybridization of the LLs induces new optical selection rules for modulo 3 in addition to the $\Delta n=\pm 1$ applied to transitions between the well-defined LLs. Moreover, the rich magnetic quantization leads to the complicated electronic excitations, mainly owing to the strong competition between the transverse Lorentz motion and the longitudinal Coulomb oscillation. There are a lot of low-frequency magneto-plasmons with a novel momentum-frequency relationship. The AA- and AB-stacked bilayer graphenes, respectively, exhibit the 2D-like plasmon and the discrete inter-LL plasmons. For multilayer graphene, the main characteristics of magneto-plasmons, existence, intensity and frequency, will be significantly affected by the stacking configuration and the layer number.

Ever since graphene has first been successfully fabricated, its unique physical properties have drawn intensive attentions of scientists that are devoted to researching new two-dimensional materials for possible applications in next-generation nanoelectronic devices and fundamental scientific research. Recently, 2D materials like MoS$_{2}$\cite{NatTech6;147,PRL105;136805} and silicene\cite{APL96;183102,PRL108;155501,PRL108;245501}, arranged within a buckled honeycomb lattice, have been synthesized and present an intrinsic bandgap\cite{NatTech6;147,PRL105;136805,PRL108;155501,PRL108;245501}. The electronic dispersion resembles that of relativistic Dirac fermions. Likewise, the generalized tight-binding model can cope with the $sp^{3}$ hybridizations in the buckled structure and handle the influence of external fields; it can be adapted to different geometric structures, including bounded, folded, curved, and stacked configurations.

\bigskip

\bigskip

\centerline {\textbf {ACKNOWLEDGMENTS}}%

\bigskip

\bigskip

\noindent
This work was supported in part by the Ministry of Science and Technology of Taiwan, the Republic of China, under Grant Nos. NSC 98-2112-M-006-013-MY4 and NSC 99-2112-M-165-001-MY3.

\centerline {\Large \textbf {References}}

$^{\ddag}$e-mail address: l2894110@mail.ncku.edu.tw

$^{\dag}$e-mail address: airegg.py90g@nctu.edu.tw

$^{*}$e-mail address: mflin@mail.ncku.edu.tw

\bigskip \vskip0.6 truecm

\noindent

\newpage

\centerline {\Large \textbf {Figure Captions}}

Figure 1. (a) Honeycomb lattice structure of monolayer graphene. Sublattices A and B are shown by black and red colors, respectively. The primitive unit cell is depicted by a gray diamond, where $\alpha$ indicates the hopping integral between two sublattices A and B and $b^{'}$ represents the C-C bond length. (b). The Brillouin zone of honeycomb lattice and some highly symmetric points. (c). The magnetically induced rectangle unit cell under a uniform magnetic field $\textbf{B}=B_{0}\hat{z}$, perpendicular to the graphene plane.

Figure 2. Geometric structure of AA-stacked graphene. The interlayer atomic interactions are illustrated in the right panel.

Figure 3. Geometric structure of AB-stacked graphene. The interlayer atomic interactions are illustrated in the right panel.

Figure 4. Geometric structure of ABC-stacked graphene. The interlayer atomic interactions are illustrated in the right panel.

Figure 5.  Band structure of monolayer graphene along $\Gamma\rightarrow K\rightarrow M$ direction indicated in Fig. 1(b).

Figure 6. (a). Landau levels of monolayer graphene at $B_{0}=20$ T. The quantum numbers $n^{c,v}=0,1,2,3...$ are counted from the Fermi level, $E_{F}=0$, to higher conduction and valence energies. (b) and (c). The envelope functions of two sublattices A and B are plotted with respect to the Landau levels shown in (a).

Figure 7. Energy dependence of Landau level on the square root of (a) magnetic field strength and (b) quantum number.

Figure 8. Band structure of monolayer graphene under a modulated magnetic field with the modulation field strength $B_{M}=20$ T and the period length $R_{M}=500$. The center of the oscillating Landau subbands is localized at $k_{y}=2/3$. Red lines indicate the energies of the Landau levels under a uniform field of 20 T.

Figure 9. (a) Quasi-Landau levels of flat graphene ribbon. Oscillating Landau subbands of (b) curved graphene ribbon and (c) carbon nanotube. The parameters $W$, $\theta$ and $R$ represent the width of ribbon, the arc angle of curved ribbon and the radius of tube, respectively.

Figure 10. Zero-field band structure of AA-stacked graphenes for (a) odd number of layers: monolayer and trilayer, and (b) even number of layers: bilayer and tetralayer.

Figure 11. (a) Landau levels of AA-stacked trilayer graphene under $B_{0}=20$ T. For each level, the envelop functions of sublattices $A_{o}^{1}$, $B_{o}^{1}$, $A_{o}^{2}$, $B_{o}^{2}$, $A_{o}^{3}$ and $B_{o}^{3}$ for the 1/6 localization state are shown in (b), (c), (d), (e), (f) and (g), respectively.

Figure 12. Landau-level spectrum of AA-stacked trilayer graphene. The Fermi level is plotted as a bold wiggling curve, an oscillation around the zero energy as a function of the magnetic field strength.

Figure 13. Landau-level spectrum of AA-stacked bilayer graphene. The Fermi level is plotted as a bold wriggling curve, an oscillation around the zero energy as a function of the magnetic field strength.

Figure 14. Landau-level spectrum of AA-stacked tetralayer graphene. The Fermi level is plotted as a bold wriggling curve, an oscillation around the zero energy as a function of the magnetic field strength.

Figure 15. Metal-semiconductor transitions during a variation of the magnetic field for AA-stacked (a) bilayer and (b) trilayer graphenes with/without an applied gate voltage.

Figure 16. Band structure of AB-stacked (a) bilayer and (b) trilayer graphens in the absence of external fields. The inserts of (a) and (b) show the zoomed-in view near the Fermi level.

Figure 17. (a) Landau levels of AB-stacked trilayer graphene under $B_{0}=20$ T. For each level, the envelop functions of sublattices $A_{o}^{1}$, $B_{o}^{1}$, $A_{o}^{2}$, $B_{o}^{2}$, $A_{o}^{3}$ and $B_{o}^{3}$ for the 1/6 localization state are shown in (b), (c), (d), (e), (f) and (g); those for 2/6 localization state are shown in (h), (i), (j), (k), (l) and (m), respectively.

Figure 18. Landau-level spectrum of AB-stacked trilayer graphene.

Figure 19. Landau-level spectrum of AB-stacked bilayer graphene.

Figure 20. Landau-level spectrum of AB-stacked tetralayer graphene.

Figure 21. Band structure of ABC-stacked (a) trilayer and (b) tetralayer graphenes in the absence of external fields. The inserts of (a) and (b) show the zoomed-in view near the Fermi level, where the crossing points are the Dirac points.

Figure 22. (a) Landau levels of ABC-stacked trilayer graphene under $B_{0}=20$ T. For each level, the odd-indexed envelop functions of sublattices $A_{o}^{1}$, $B_{o}^{1}$, $A_{o}^{2}$, $B_{o}^{2}$, $A_{o}^{3}$ and $B_{o}^{3}$  for the 2/6 localization state are shown in (b), (c), (d), (e), (f) and (g), respectively.

Figure 23. Landau-level spectrum of ABC-stacked trilayer graphene.

Figure 24. (a) Landau-level anticrossing pattern of ABC-stacked trilayer graphene. For the states marked by circles in (a), their envelop functions of sublattices $A_{o}^{1}$, $B_{o}^{1}$, $A_{o}^{2}$, $B_{o}^{2}$, $A_{o}^{3}$ and $B_{o}^{3}$ for the 2/6 localization state are plotted in (b), (c), (d), (e), (f), and (g), respectively.

Figure 25.  Landau-level spectrum of ABC-stacked tetralayer graphene.

Figure 26.  Landau-level spectrum of ABC-stacked pentalayer graphene. The main characteristics of the first to fifth group are plotted in (a), (b), (c), and (d).

Figure 27. Energies of Dirac-point related Landau levels of ABC-stacked trilayer graphene are plotted for $B_{0}$=25 T based on their dependence on the number of stacked layers.

Figure 28. (a) Landau levels of the second group of ABC-stacked trilayer graphene are plotted based on the full tight-binding model (black) and the minimum model (red). Red cross symbols indicate the highest Landau levels which are fully occupied under the critical fields $B_{s}$. (b) Linear relationship between $B_{s}$ and $n_{s}$ is plotted, where $n_{s}$ appears with a constant period of $1/B_{0}$.

Figure 29. Density of states of Landau levels for (a) AA-, (b) AB-, and (c) ABC-stacked trilayer graphenes.

\newpage
\begin{figure}[tbp]
\par
\begin{center}
\leavevmode
\includegraphics[width=0.8\linewidth]{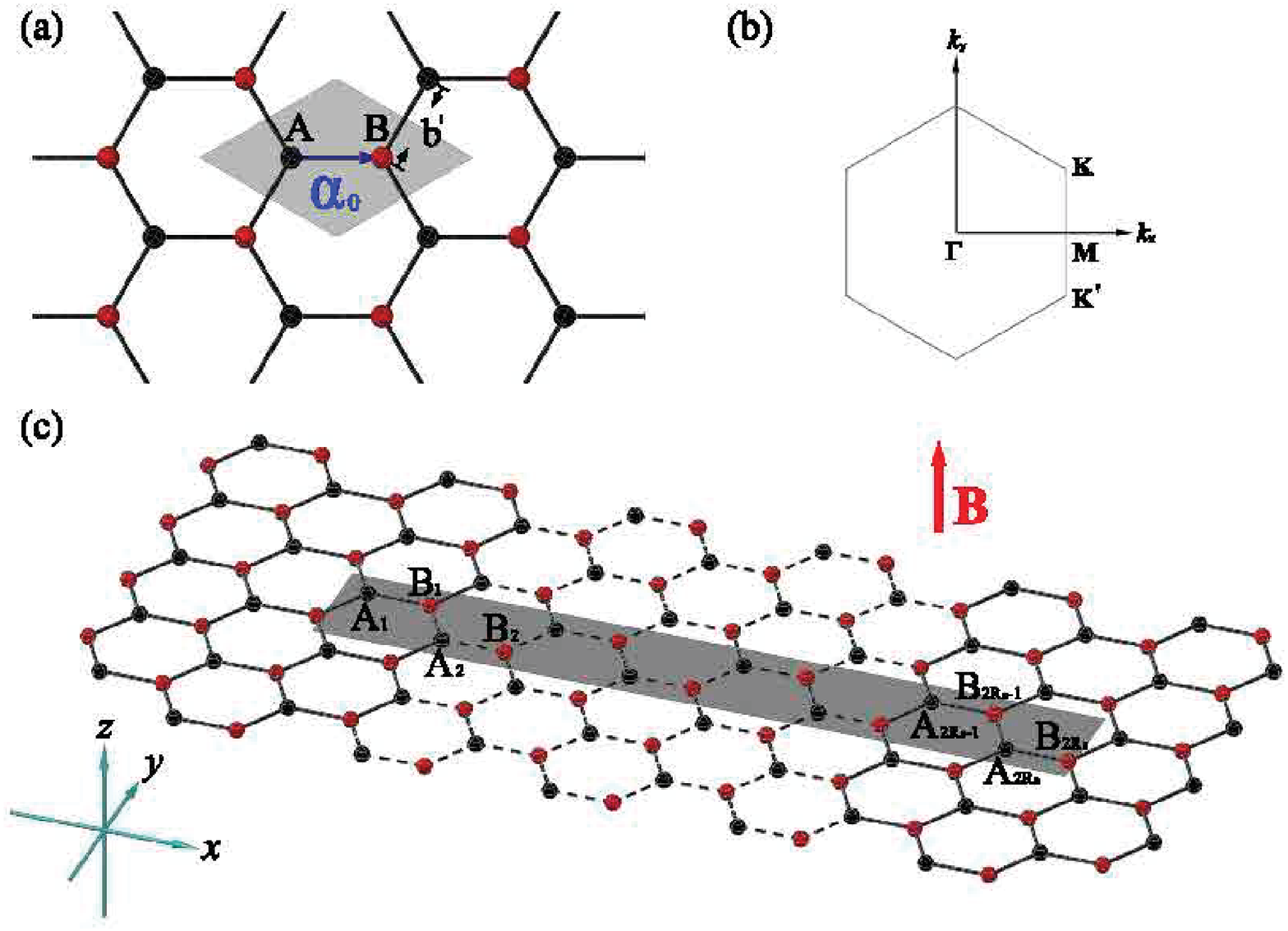}
\end{center}
\par
\textbf{Figure 1}
\end{figure}
\begin{figure}[tbp]
\par
\begin{center}
\leavevmode
\includegraphics[width=0.7\linewidth]{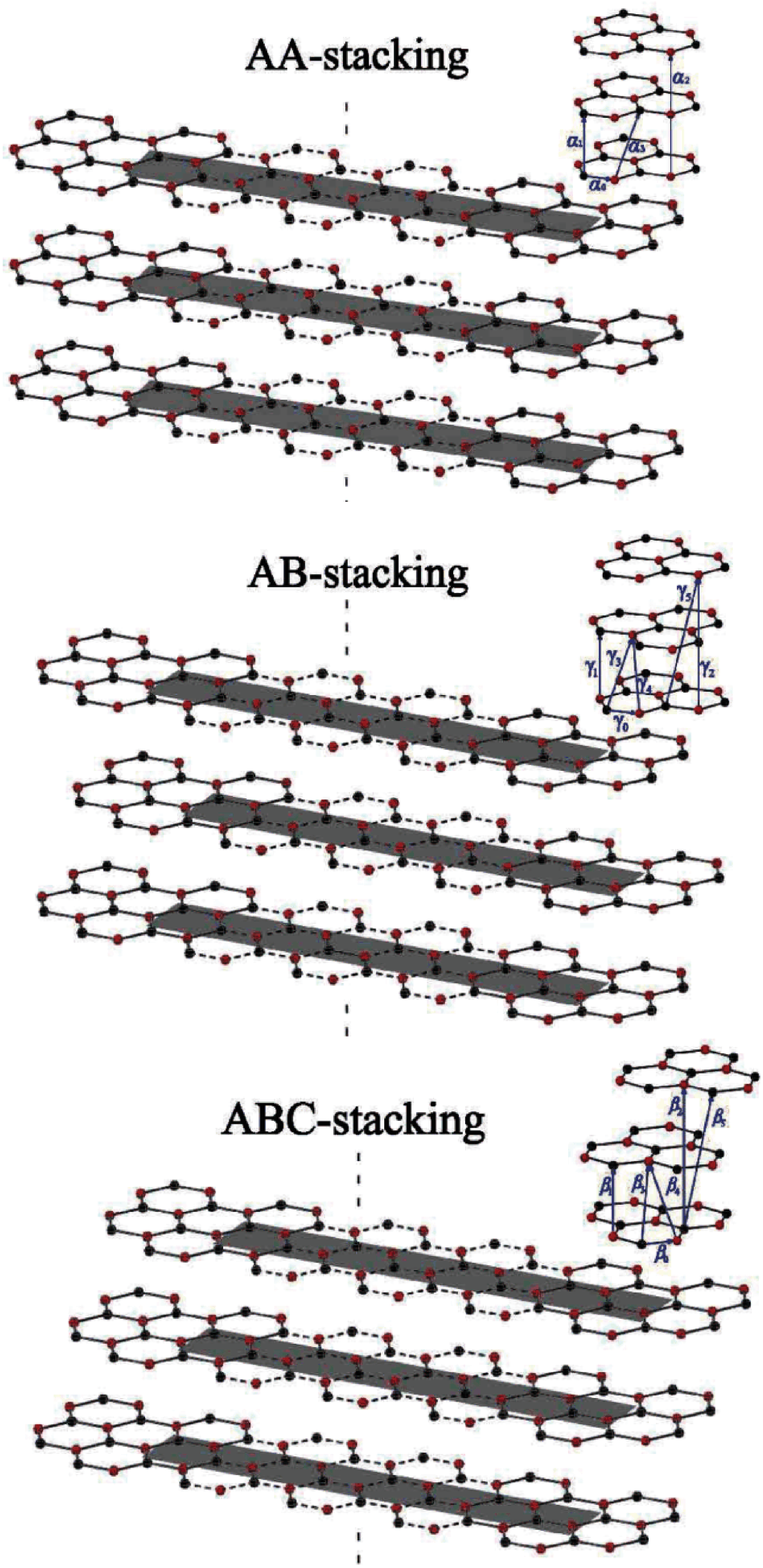}
\end{center}
\par
\textbf{Figures 2, 3 and 4}
\end{figure}
\begin{figure}[tbp]
\par
\begin{center}
\leavevmode
\includegraphics[width=0.8\linewidth]{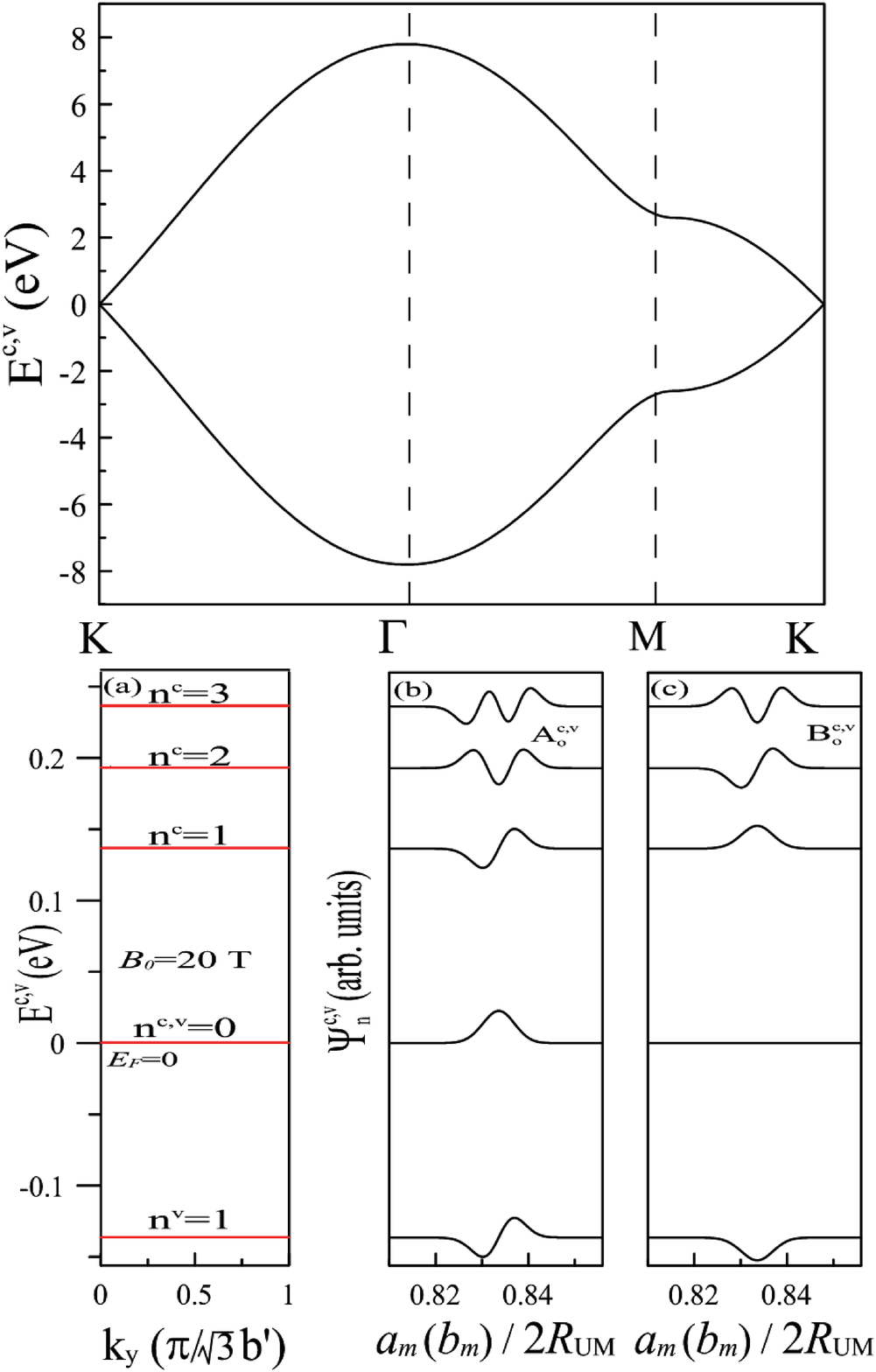}
\end{center}
\par
\textbf{Figures 5 and 6}
\end{figure}
\begin{figure}[tbp]
\par
\begin{center}
\leavevmode
\includegraphics[width=0.8\linewidth]{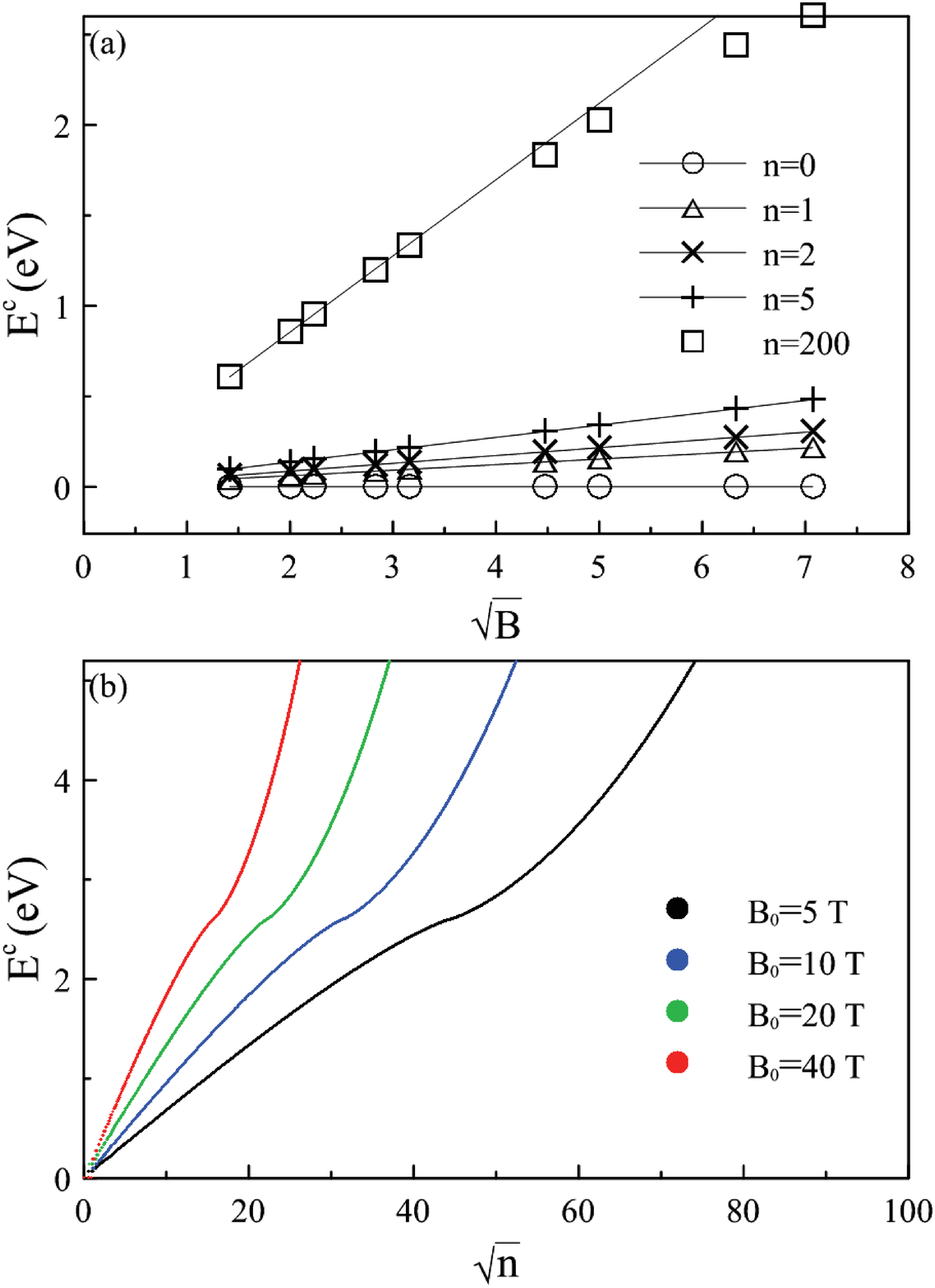}
\end{center}
\par
\textbf{Figure 7}
\end{figure}
\begin{figure}[tbp]
\par
\begin{center}
\leavevmode
\includegraphics[width=0.8\linewidth]{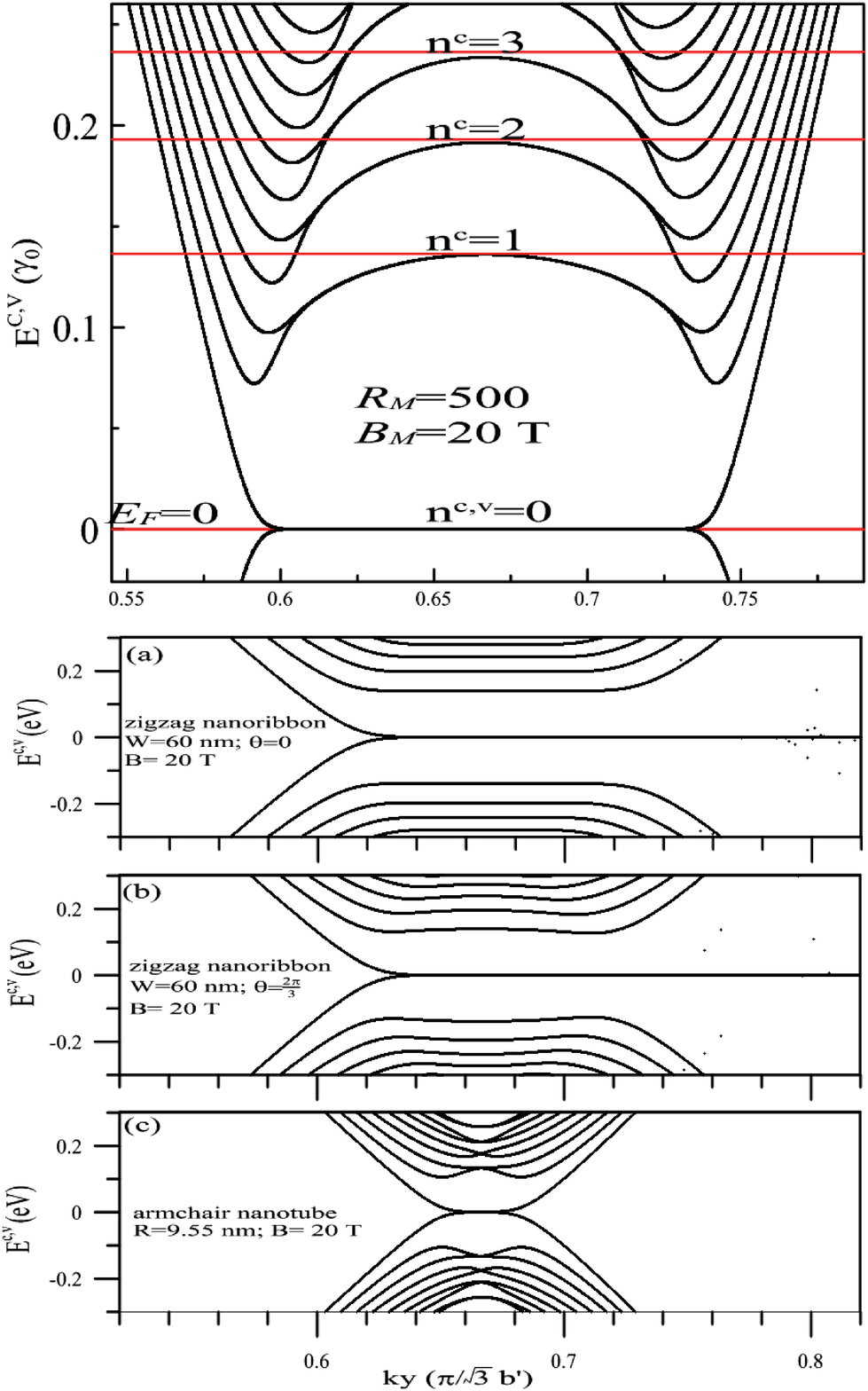}
\end{center}
\par
\textbf{Figures 8 and 9}
\end{figure}
\begin{figure}[tbp]
\par
\begin{center}
\leavevmode
\includegraphics[width=0.8\linewidth]{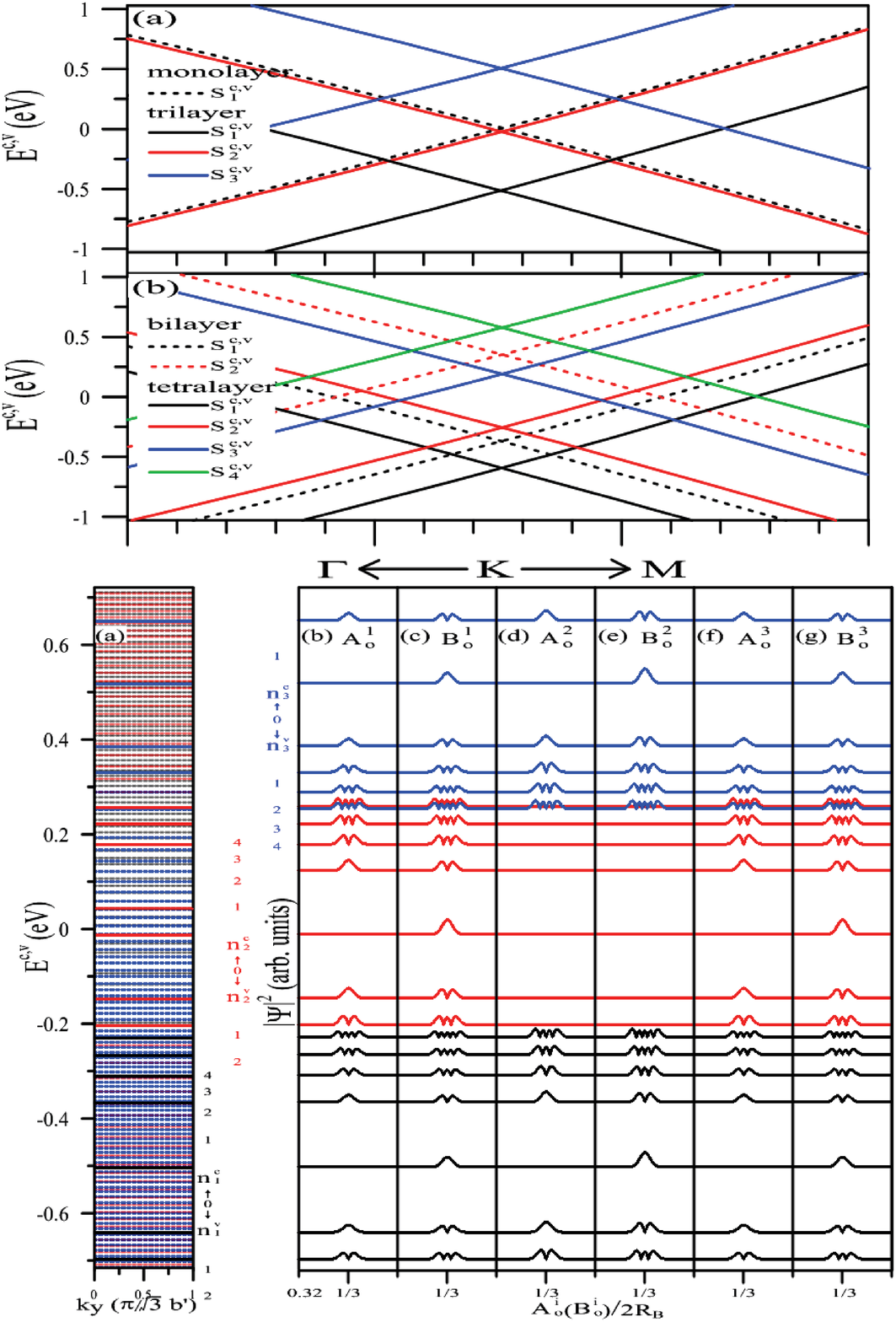}
\end{center}
\par
\textbf{Figures 10 and 11}
\end{figure}
\begin{figure}[tbp]
\par
\begin{center}
\leavevmode
\includegraphics[width=0.8\linewidth]{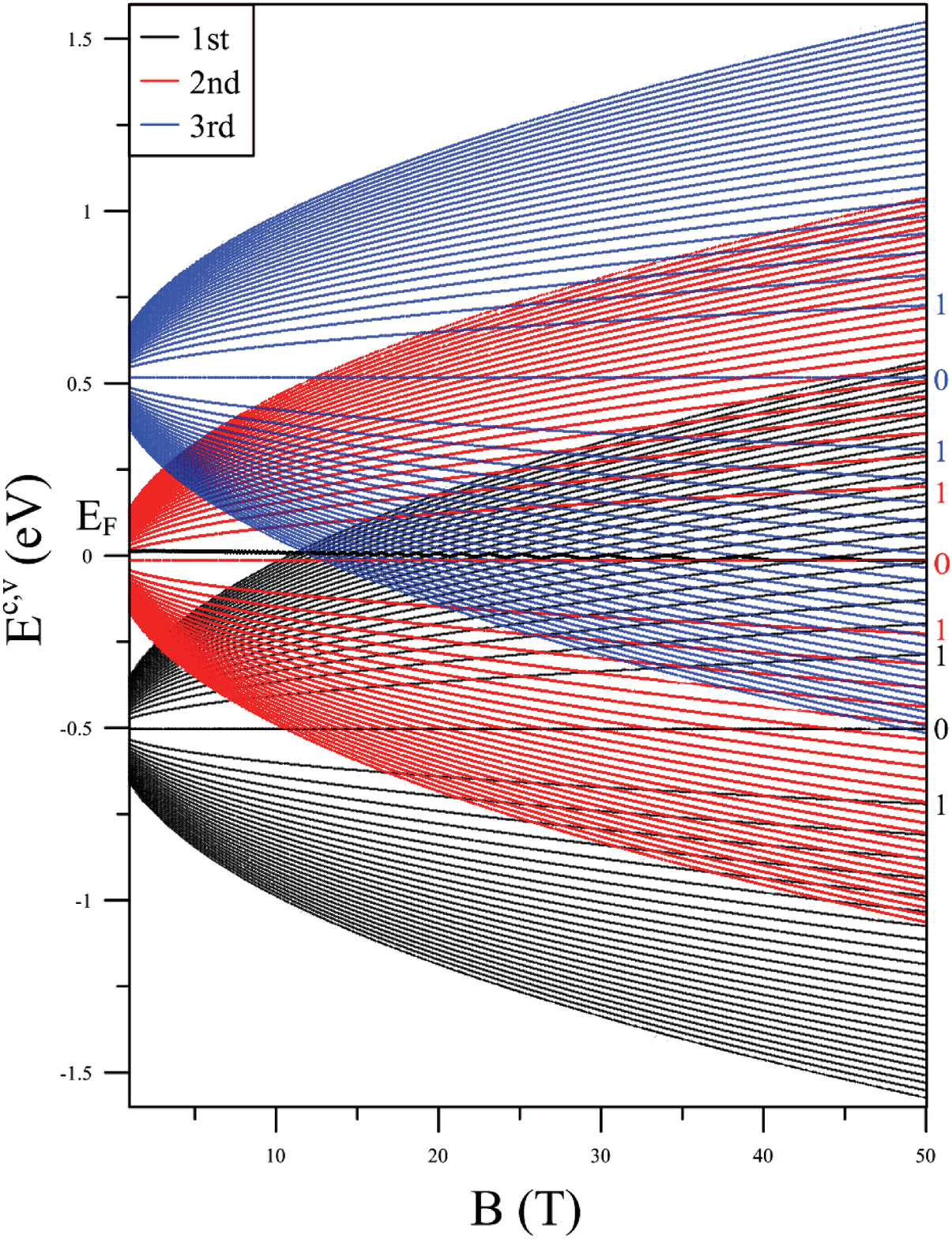}
\end{center}
\par
\textbf{Figure 12}
\end{figure}
\begin{figure}[tbp]
\par
\begin{center}
\leavevmode
\includegraphics[width=0.8\linewidth]{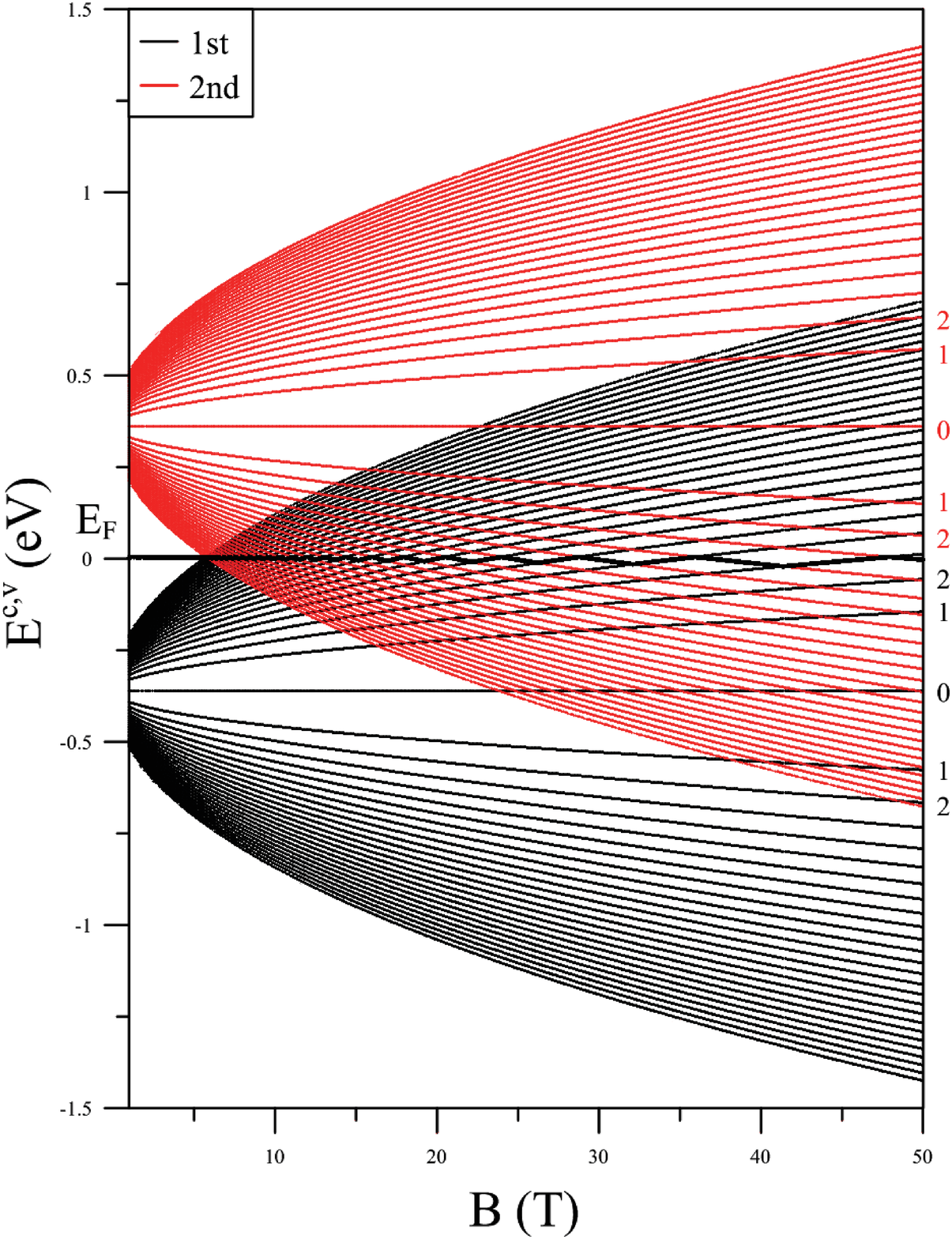}
\end{center}
\par
\textbf{Figure 13}
\end{figure}
\begin{figure}[tbp]
\par
\begin{center}
\leavevmode
\includegraphics[width=0.8\linewidth]{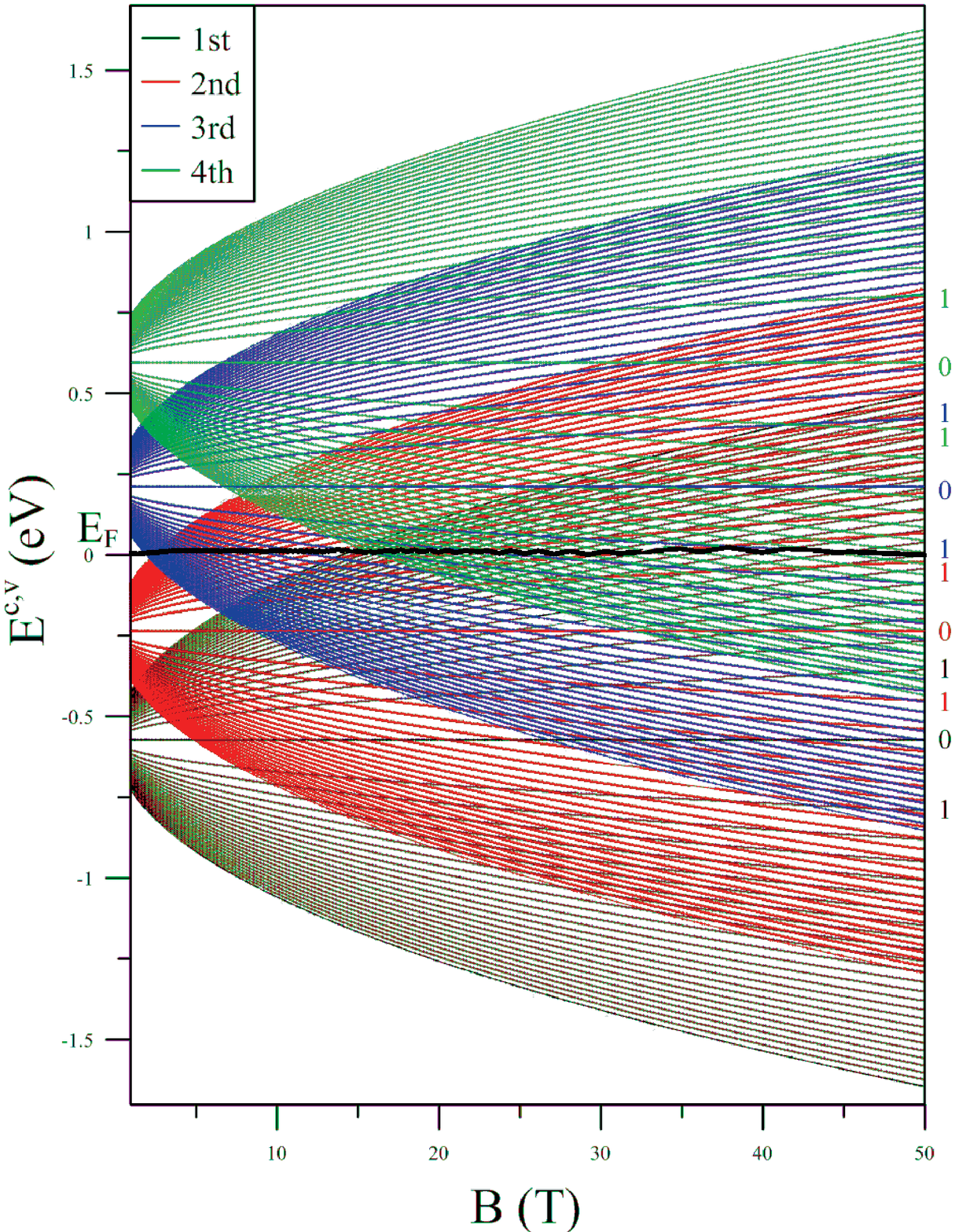}
\end{center}
\par
\textbf{Figure 14}
\end{figure}
\begin{figure}[tbp]
\par
\begin{center}
\leavevmode
\includegraphics[width=0.8\linewidth]{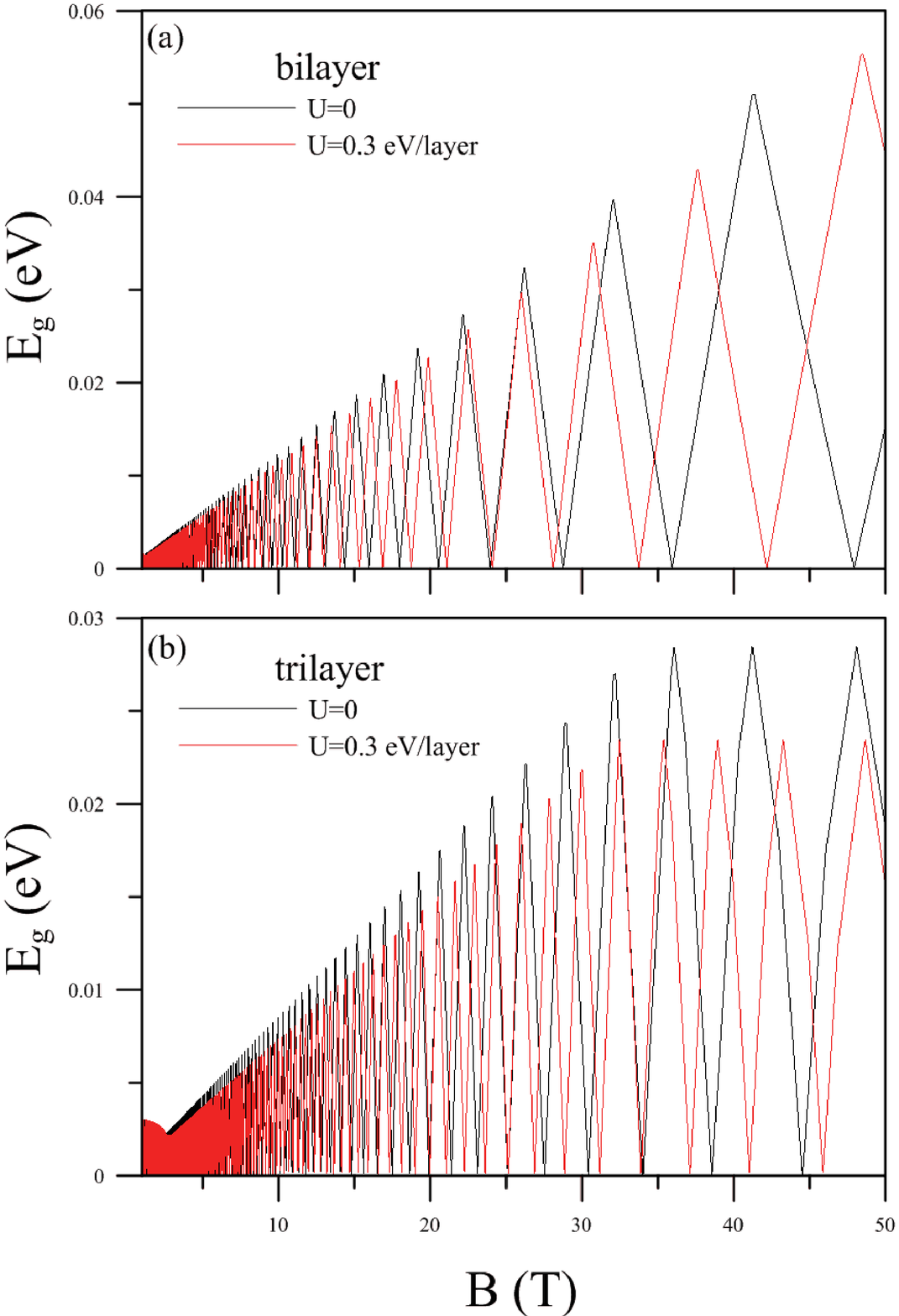}
\end{center}
\par
\textbf{Figure 15}
\end{figure}
\begin{figure}[tbp]
\par
\begin{center}
\leavevmode
\includegraphics[width=0.8\linewidth]{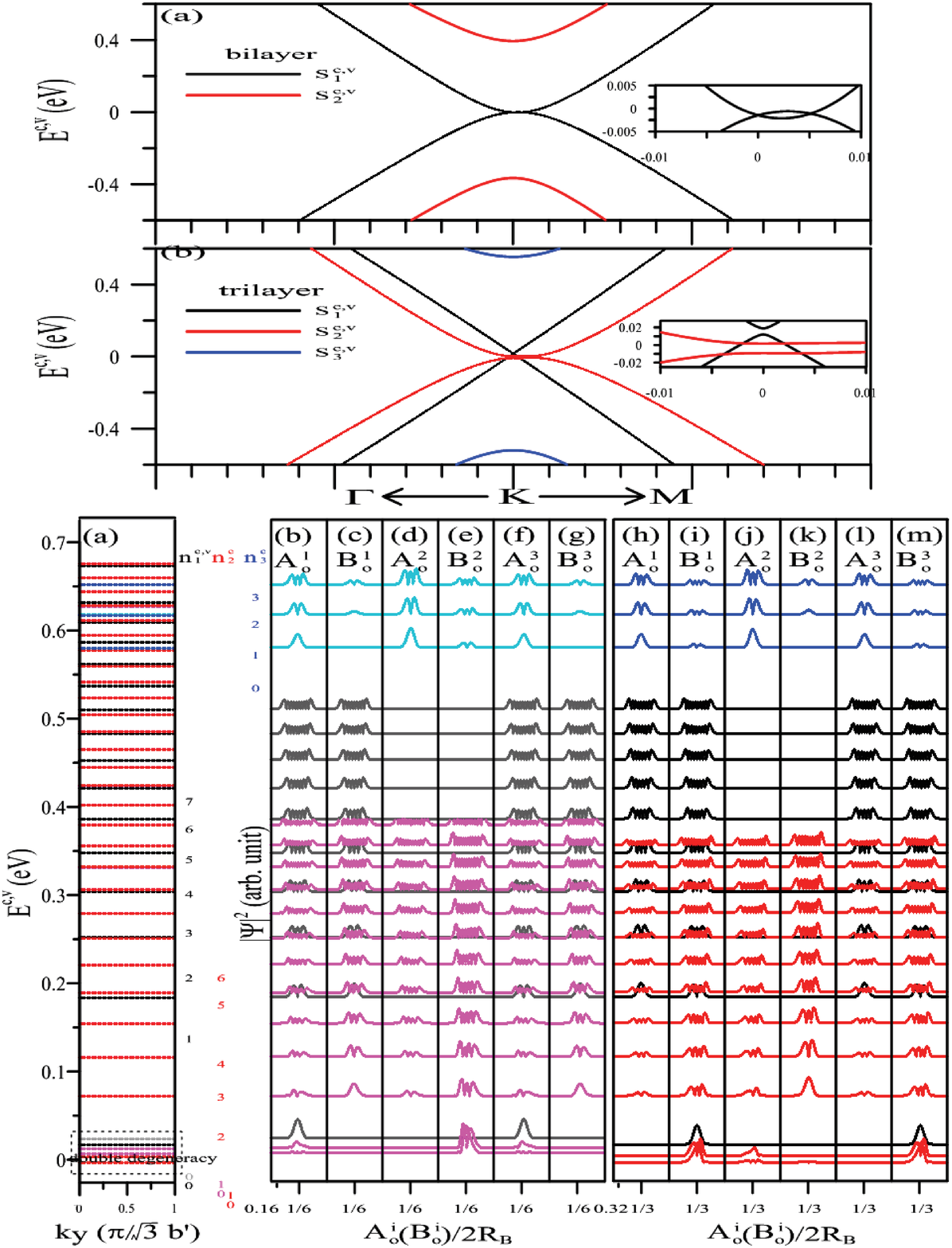}
\end{center}
\par
\textbf{Figures 16 and 17}
\end{figure}
\begin{figure}[tbp]
\par
\begin{center}
\leavevmode
\includegraphics[width=0.8\linewidth]{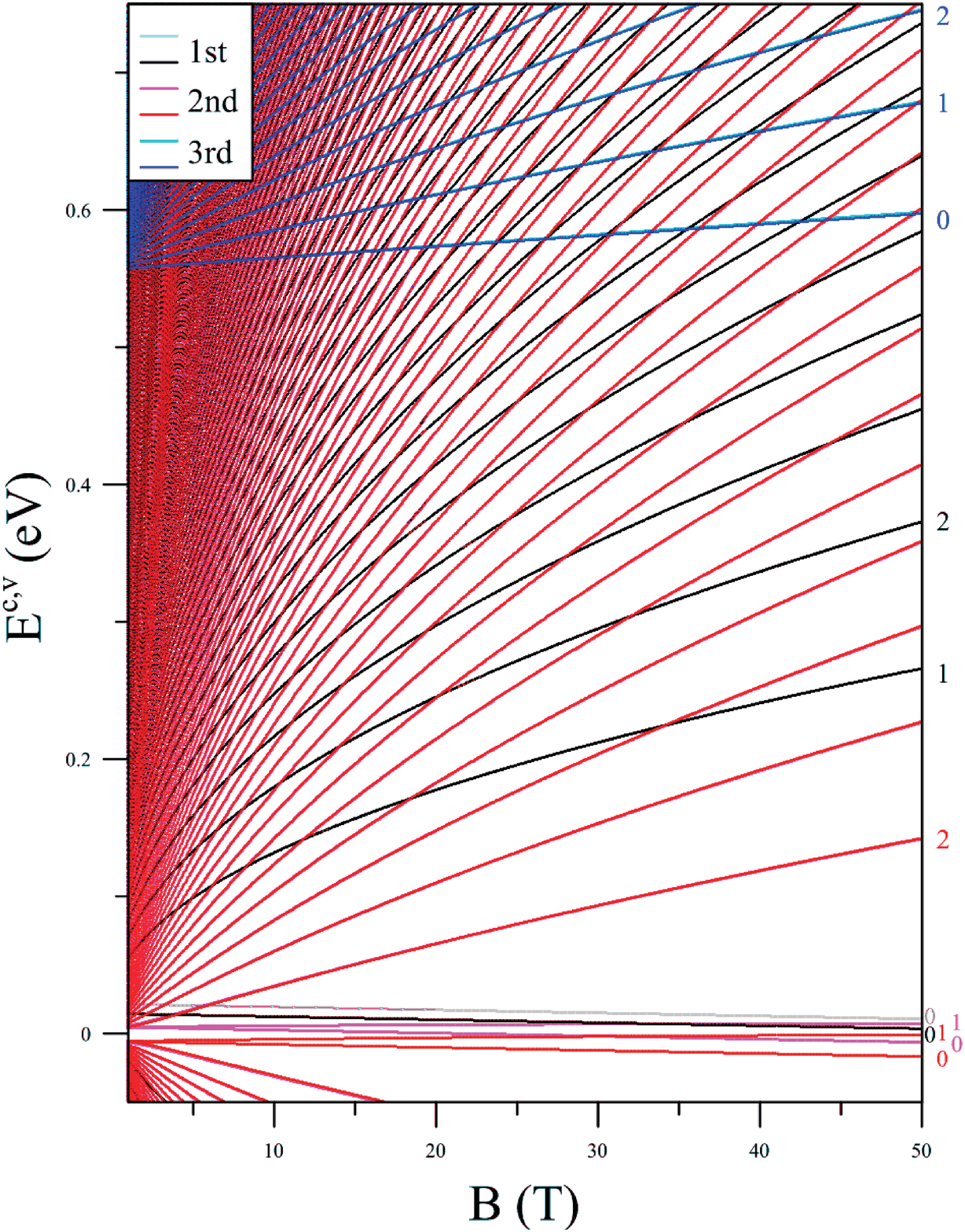}
\end{center}
\par
\textbf{Figure 18}
\end{figure}
\begin{figure}[tbp]
\par
\begin{center}
\leavevmode
\includegraphics[width=0.8\linewidth]{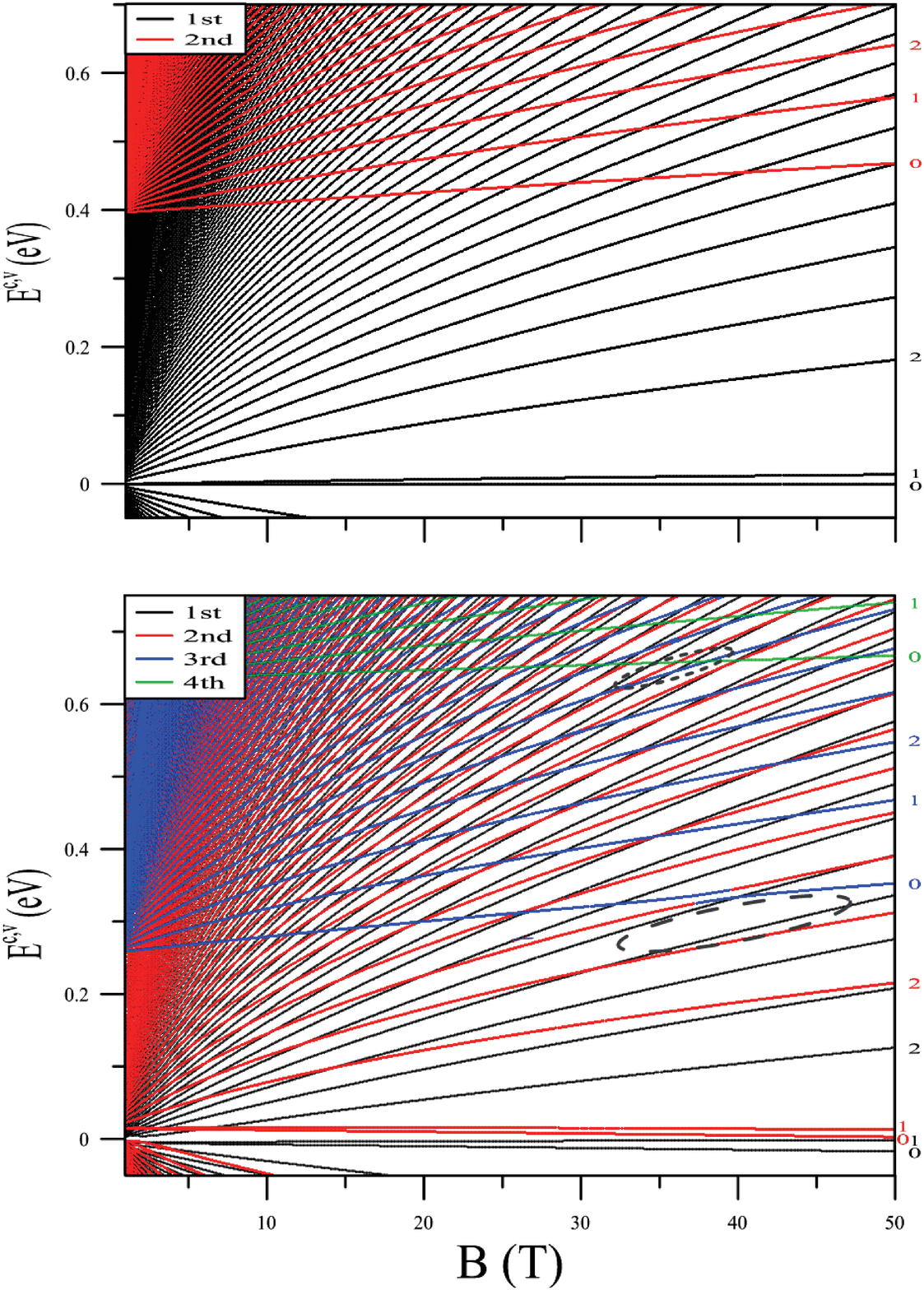}
\end{center}
\par
\textbf{Figures 19 and 20}
\end{figure}
\begin{figure}[tbp]
\par
\begin{center}
\leavevmode
\includegraphics[width=0.8\linewidth]{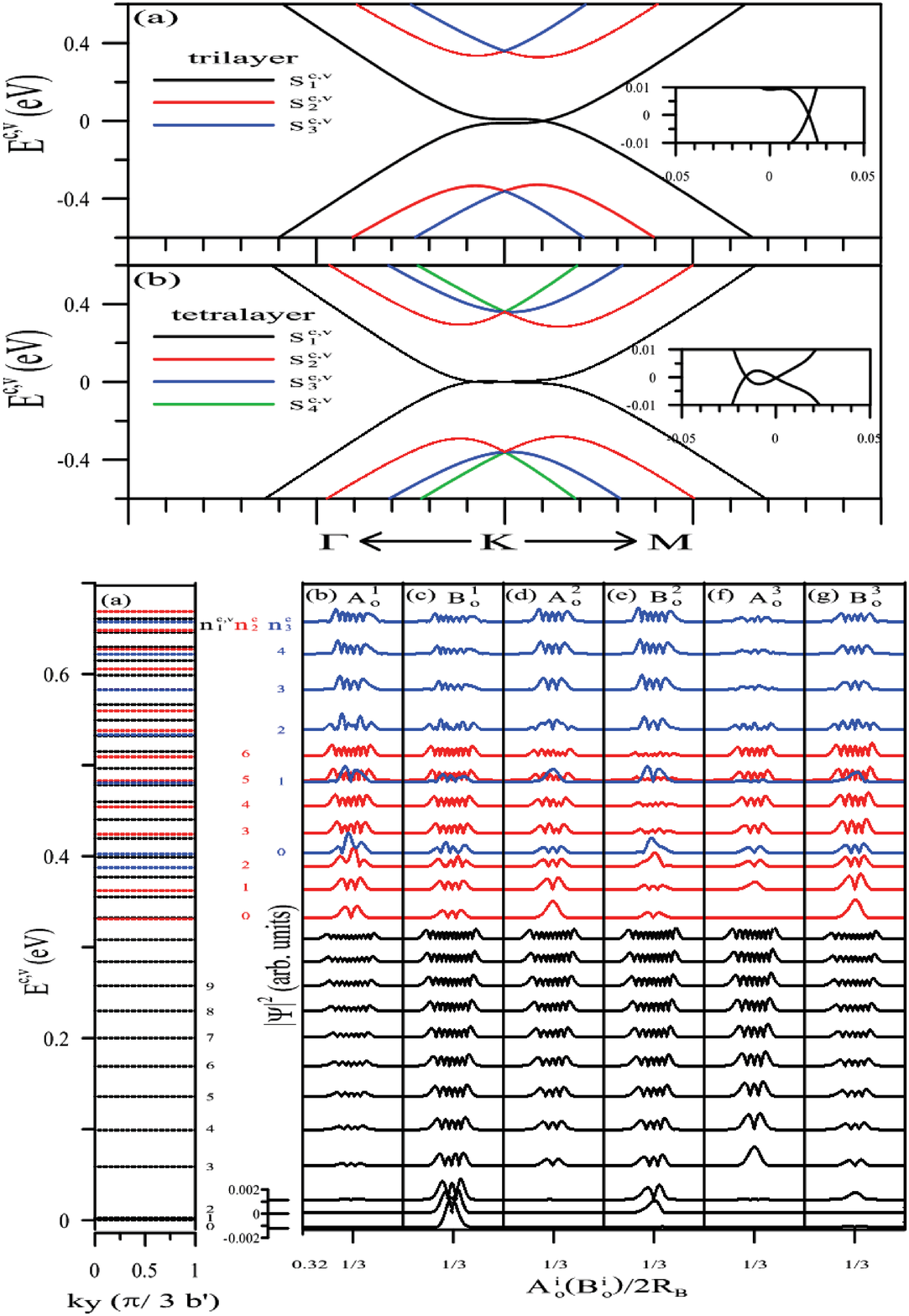}
\end{center}
\par
\textbf{Figures 21 and 22}
\end{figure}
\begin{figure}[tbp]
\par
\begin{center}
\leavevmode
\includegraphics[width=0.8\linewidth]{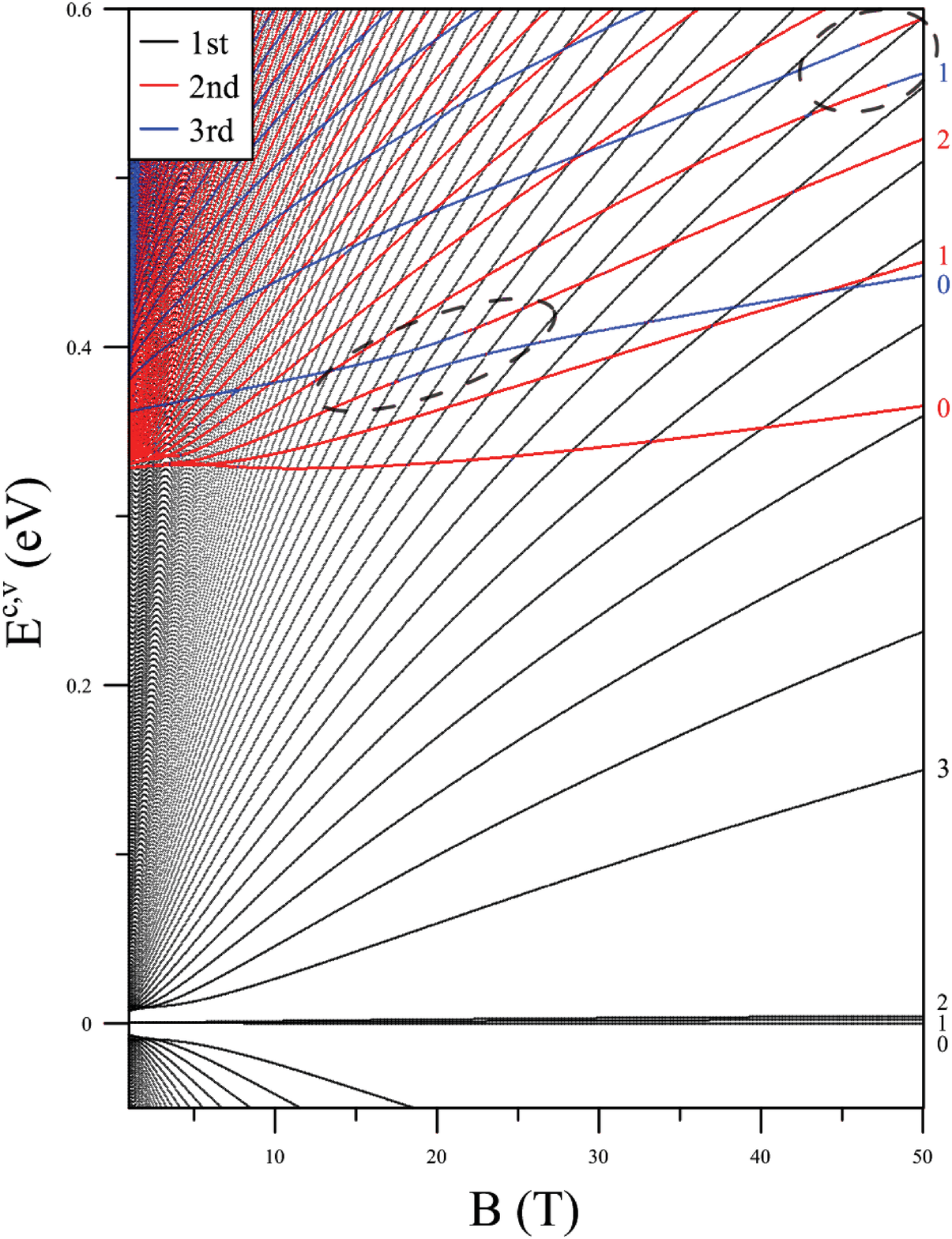}
\end{center}
\par
\textbf{Figure 23}
\end{figure}
\begin{figure}[tbp]
\par
\begin{center}
\leavevmode
\includegraphics[width=0.8\linewidth]{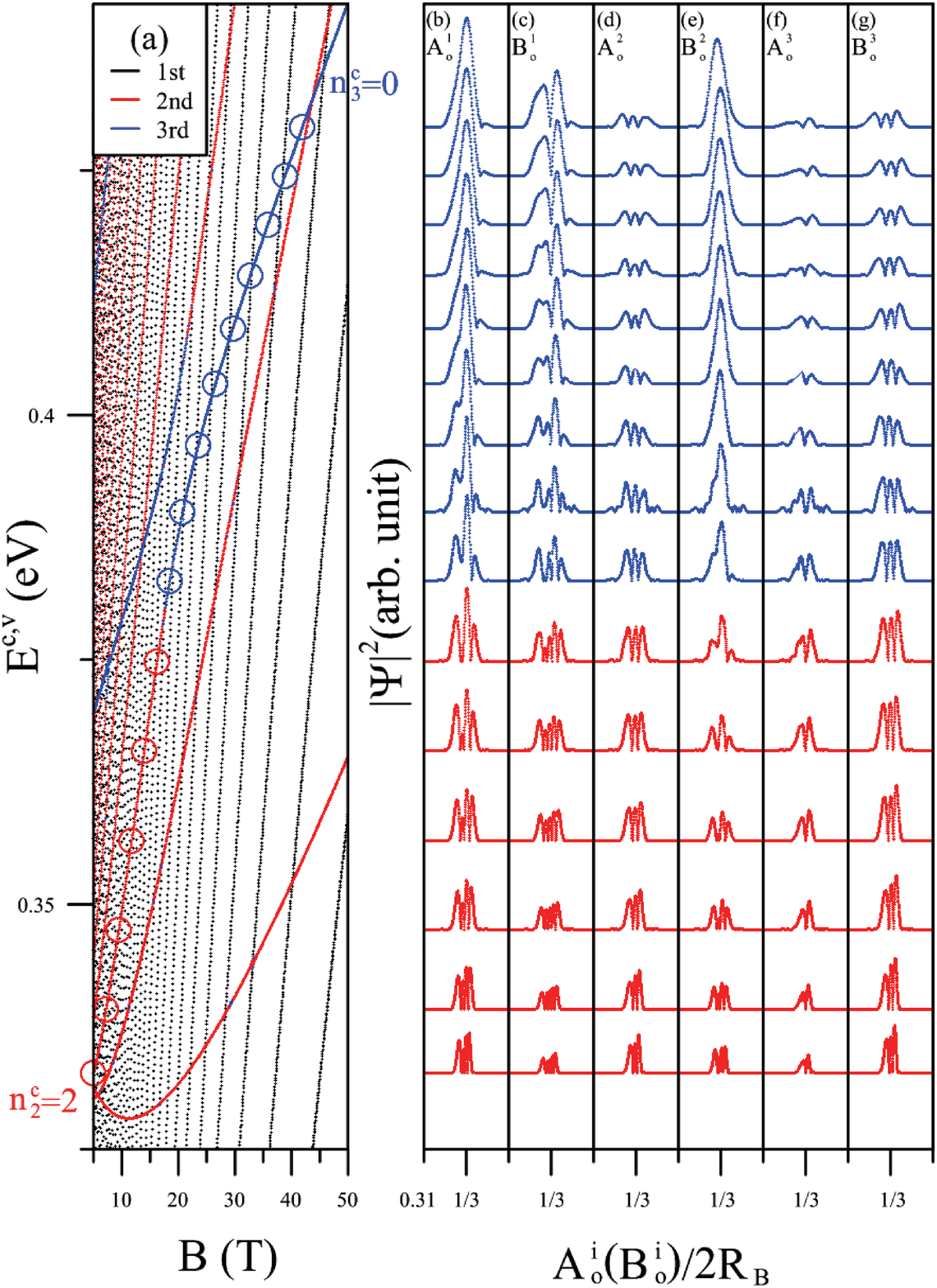}
\end{center}
\par
\textbf{Figure 24}
\end{figure}
\begin{figure}[tbp]
\par
\begin{center}
\leavevmode
\includegraphics[width=0.8\linewidth]{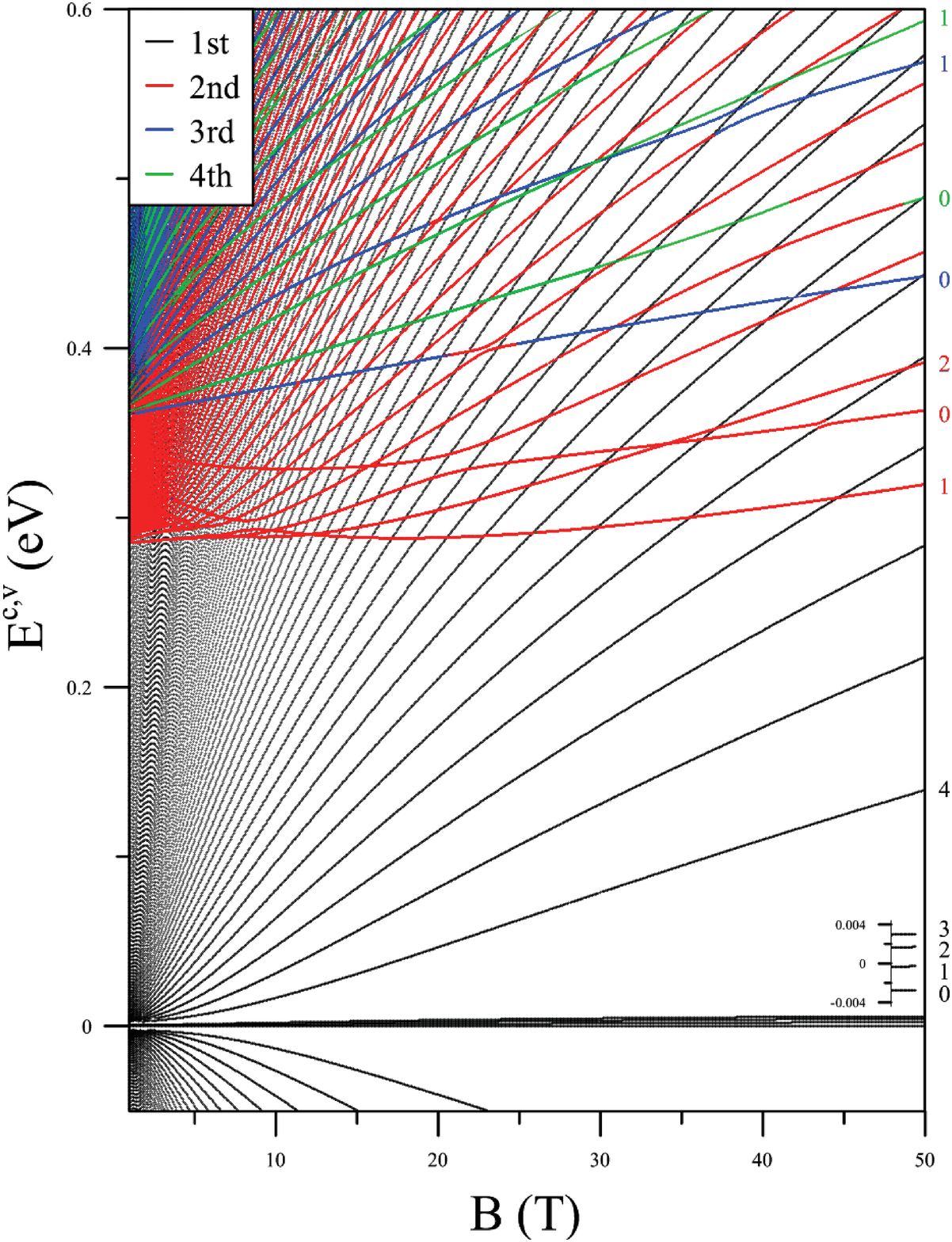}
\end{center}
\par
\textbf{Figure 25}
\end{figure}
\begin{figure}[tbp]
\par
\begin{center}
\leavevmode
\includegraphics[width=0.8\linewidth]{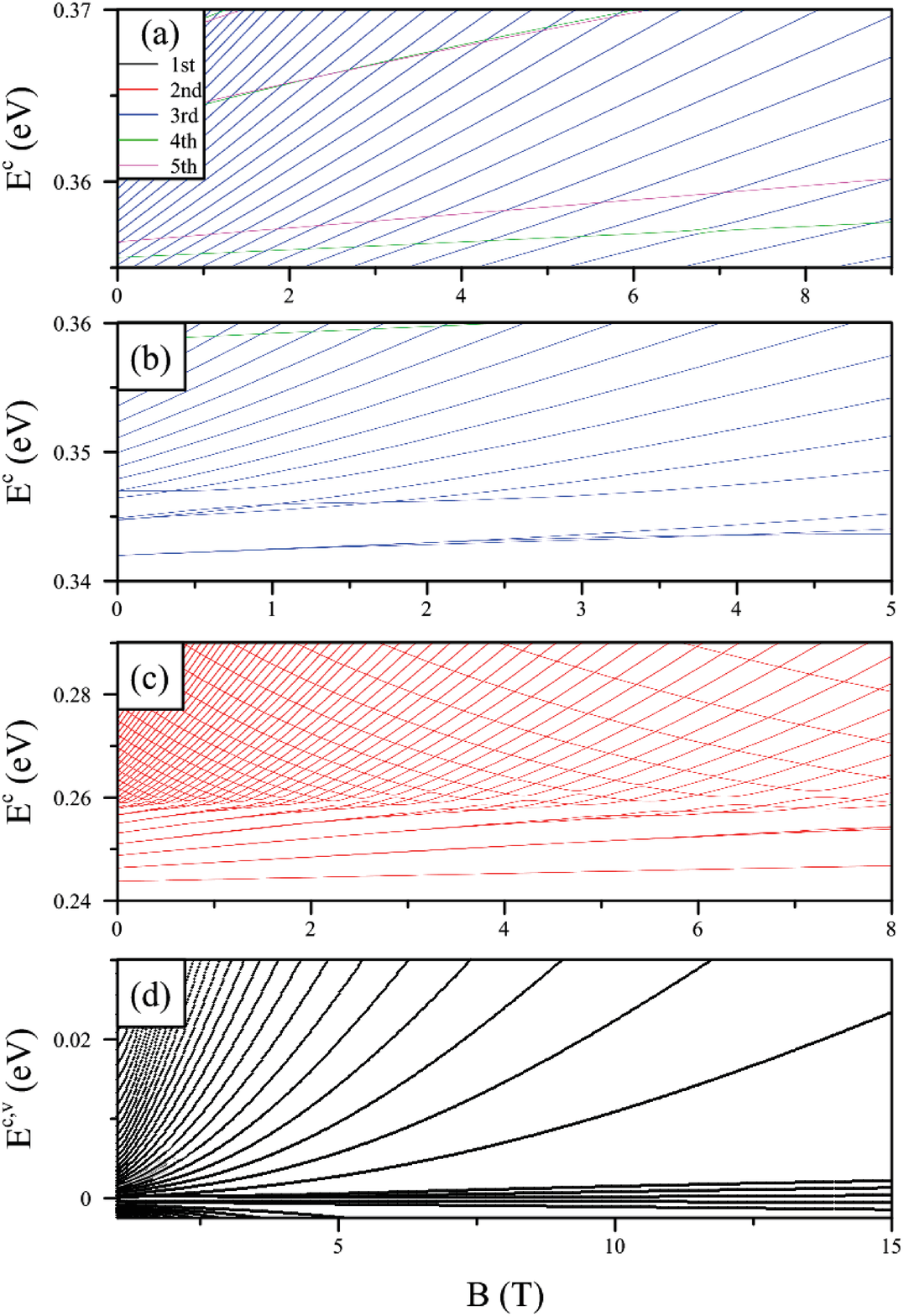}
\end{center}
\par
\textbf{Figure 26}
\end{figure}

\clearpage

\begin{figure}[tbp]
\par
\begin{center}
\leavevmode
\includegraphics[width=0.8\linewidth]{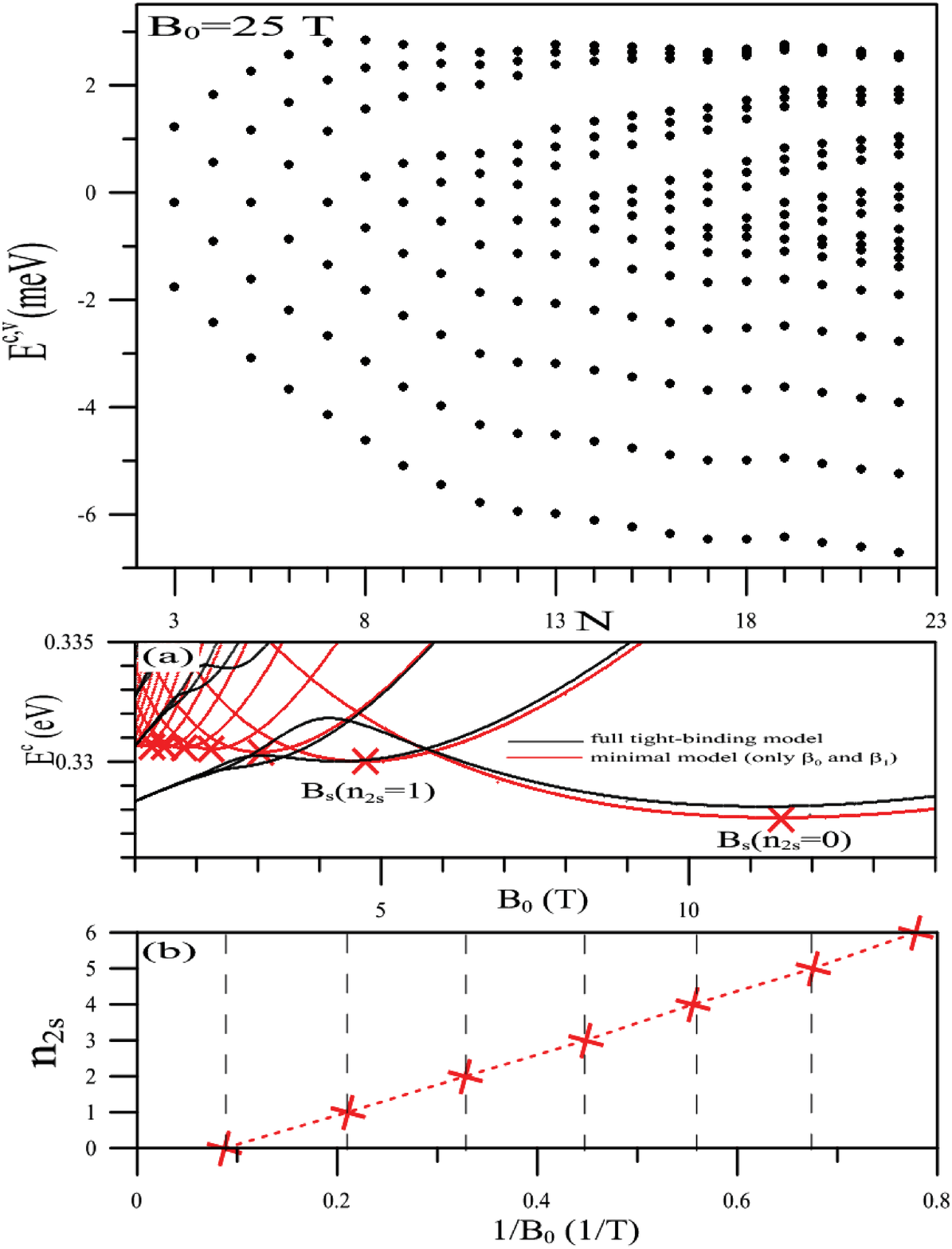}
\end{center}
\par
\textbf{Figures 27 and 28}
\end{figure}

\begin{figure}[tbp]
\par
\begin{center}
\leavevmode
\includegraphics[width=0.8\linewidth]{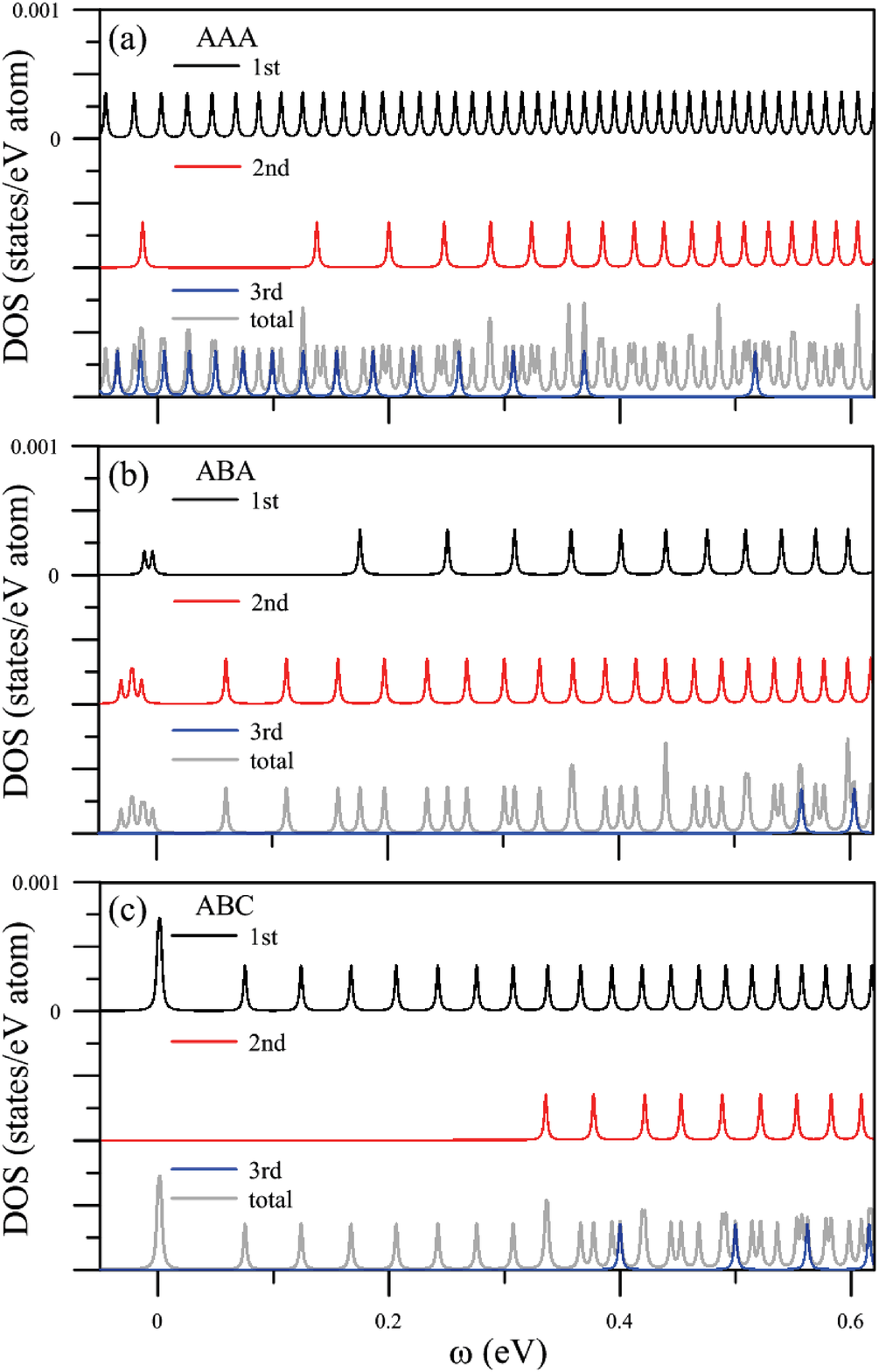}
\end{center}
\par
\textbf{Figure 29}
\end{figure}

\end{document}